\documentclass[aps,a4,twocolumn,superscriptaddress,preprintnumbers,nofootinbib]{revtex4}
\pdfoutput=1

\usepackage{pslatex}
\usepackage[pdftex]{graphicx}
\usepackage{psfrag}
\usepackage{epsfig}
\usepackage{color}
\usepackage{cancel}
\usepackage{slashed}
\usepackage{amssymb}
\usepackage{amsmath}
\usepackage{hyperref}
\usepackage{enumerate}
\usepackage{multirow}
\usepackage{lipsum}
\usepackage[normalem]{ulem}

\usepackage{orcidlink}

\definecolor{vdrgreen}{rgb}{0.0, 0.6, 0.0}

\begin{document}

\title{Axionlike particle searches in short-baseline liquid
  scintillator neutrino detectors}%

\author{D. Aristizabal Sierra\orcidlink{}}%
\email{daristizabal@uliege.be}%
\affiliation{Universidad T\'ecnica Federico Santa Mar\'{i}a -
  Departamento de F\'{i}sica\\Casilla 110-V, Avda. Espa\~na 1680,
  Valpara\'{i}so, Chile}%
\author{L. Duque\orcidlink{}}%
\email{laura.duque@cinvestav.mx}%
\affiliation{Departamento de F\'{\i}sica, Centro de Investigaci\'on y
  de Estudios Avanzados del IPN,\\ Apartado Postal 14-740 07000
  Mexico, Distrito Federal, Mexico}%
\author{O. Miranda\orcidlink{0000-0002-0310-7060}} \email{omar.miranda@cinvestav.mx}
\affiliation{Departamento de F\'{\i}sica, Centro de Investigaci\'on y
  de Estudios Avanzados del IPN,\\ Apartado Postal 14-740 07000
  Mexico, Distrito Federal, Mexico} \author{H. Nunokawa\orcidlink{0000-0002-3369-0840}
}%
\email{nunokawa@puc-rio.br}%
\affiliation{Pont\'{i}ficia Universidade Cat\'{o}lica do Rio de Janeiro
  (PUC-Rio) Rua Marques de Sao Vicente, 225, C.P. 38071, 22452-970 Rio
  de Janeiro, RJ, Brazil}%
\begin{abstract}
  Short-baseline reactor neutrino experiments using large organic
  liquid scintillator detectors provide an experimentally rich
  environment for precise neutrino physics. Neutrino detection is done
  through inverse beta decay and relies on prompt and delayed signals,
  which enable powerful background discrimination. In addition to
  their neutrino program, they offer an ideal experimental environment
  for other physics searches.  Here we discuss the case of axionlike
  particles (ALPs) produced by either Primakoff-like or Compton-like
  processes. Their detection relies on the corresponding inverse
  processes, axio-electric absorption and ALP decays to photon or
  electron pairs. Assuming experimental parameter values broadly
  representative of JUNO-TAO or CLOUD we show that scattering ALP
  processes involve a prompt photon signal component followed by a
  delayed photon signal about 5 ns after. We point out that if these
  signals can be resolved, this might allow for efficient ALP signal
  discrimination against radioactive background. Likewise, coincident
  events from ALP decays and scintillation light followed by Auger
  electrons from axio-electric absorption might as well allow for
  background discrimination. We determine sensitivities in both the
  nuclear and electron channels using as a benchmark case an
  experimental setup resembling that of the CLOUD or JUNO-TAO
  detectors. Our findings demonstrate that with efficient background
  discrimination these type of technology has the capability to test
  regions in parameter space not yet explored. In particular, the
  cosmological triangle can be fully tested and regions of MeV ALP
  masses with ALP-electron couplings of the order of $10^{-8}$ can be
  entirely explored.
\end{abstract}
\maketitle
\tableofcontents
\section{Introduction}
\label{sec:intro}
Axionlike particles (ALPs) are pseudoscalar Nambu-Goldstone bosons
that arise from the spontaneous breaking of a global symmetry. They
acquire their masses either from small explicitly symmetry-breaking
terms or through quantum effects. Examples based on $U(1)$ global
symmetries include QCD axions
\cite{Peccei:1977hh,Weinberg:1977ma,Wilczek:1977pj} and Majoron models
\cite{Chikashige:1980ui}. Models involving secluded sectors, related
with the origin of dark matter, are as well good examples (see
e.g. Ref. \cite{Batell:2009yf}). Although these states are light and
feebly coupled to Standard Model (SM) fields, they have been subject
to intense experimental searches over more than four decades.

ALP scenarios are a general class of models where coupling and mass
are not interrelated, in contrast to QCD axion models where couplings
are suppressed by the symmetry-breaking scale and thus correlated with
the axion mass \cite{GrillidiCortona:2015jxo,Srednicki:1985xd}. Their
couplings and masses  ($m_a$) cover wider regions of parameter space and
therefore, are testable in a larger class of experiments. In low-mass
regions, $m_a\subset [10^{-9},10^{-4}]\,$eV,
searches have been
carried out in light shining through walls experiments---which rely on
photon-axion conversion in external electric or magnetic fields
\cite{Ehret:2010mh,Betz:2013dza,OSQAR:2015qdv}---photon polarization
experiments that leverage on light dichroism and birefringence effects
\cite{Ruoso:1992nx,Cameron:1993mr}, and fifth force searches including
torsion balance tests of the gravitational inverse square law
\cite{Kapner:2006si}.

Soon after the work of Peccei and Quinn and the recognition that such
a mechanism implied the existence of a new particle
\cite{Peccei:1977hh,Weinberg:1977ma}, searches using nuclear isotopes
de-excitations and reactor facilities were done
\cite{Zehnder:1981qn,Cavaignac:1982ek,Lehmann:1982bp,Avignone:1986vm}. These
measurements were followed up by measurements at TEXONO about 20 years
ago \cite{TEXONO:2006spf}. Given the nature of these experiments, they
are relevant in the low-MeV mass region, in contrast to those
mentioned in the previous paragraph.

In this region, astrophysical and cosmological criteria have also
played an important role. Production of these states is possible in
stellar as well as in supernova environments and at early cosmological
times.  Thus, if produced at sufficiently high rates, they can
potentially disrupt the observed dynamics of such environments
\cite{Raffelt:2006cw,Lucente:2020whw,Carenza:2020zil,Ayala:2014pea,
  Millea:2015qra,Cadamuro:2011fd,Depta:2020wmr}. Relevant in this
region as well are invisible heavy quarkonium decays, along with
beam-dump searches and collider experiments
\cite{CrystalBall:1990xec,BaBar:2008aby,Bjorken:1988as,Dobrich:2017gcm,
  Riordan:1987aw,Hearty:1989pq}. The non-observation of deviations
above expectations has placed constraints in several spots in
parameter space, but in some cases, with a few caveats. In particular,
in those spots where astrophysical and cosmological bounds apply
\cite{Jaeckel:2006xm}.

During the last decade or so there has been an increasing interest in
ALP searches in different regions of parameter space. The Axion
Resonant InterAction DetectioN Experiment (ARIADNE), leveraging on
nuclear magnetic resonance techniques, aims at testing QCD axion
models in the $10^{-4}-10^{-2}\;$eV region \cite{ARIADNE:2017tdd}. The
International Axion Observatory (IAXO), a next-generation axion
helioscope, will improve upon searches done at CAST
\cite{CAST:2015qbl} in regions of ALP masses up to the order of 1 eV
\cite{IAXO:2019mpb}. Current dark matter direct detection experiments
are making significant progress as well in the keV to MeV region and
will continue improving
\cite{XENON:2024znc,PandaX:2024sds,LZ:2023poo}. Forthcoming beam-dump
experiments, covering regions of heavier ALP masses, are expected to
increase the discovery likelihood. They include
SeaQuest~\cite{Berlin:2018pwi}, MATHUSLA~\cite{Chou:2016lxi},
CODEX-b~\cite{CODEX-b:2019jve}, FASER~\cite{Feng:2017uoz} and
SHiP~\cite{SHiP:2015vad}.

ALP searches using nuclear fission reactors is a subject that recently
has been revisited by several authors
\cite{Dent:2019ueq,AristizabalSierra:2020rom,Arias-Aragon:2023ehh}. Studies
using nuclear fusion reactors have been as well presented very
recently in Ref. \cite{Baruch:2025lbw}. Employing nuclear fission
reactors, it has attracted attention mainly because of neutrino
detectors suited for coherent elastic neutrino-nucleus scattering
measurements (see
e.g. \cite{CONNIE:2016ggr,MINER:2016igy,CONUS:2024lnu}). The same
technology used for such applications can also be used for ALP and
dark photon searches at no extra cost.  The ALP search capabilities of
a satellite experiment of the Jiangmen Underground Neutrino
Observatory (JUNO) \cite{JUNO:2015zny,JUNO:2021vlw}, the Taishan
Antineutrino Observatory (JUNO-TAO) \cite{JUNO:2020ijm}, have been
already studied in Ref. \cite{Smirnov:2021wgi}, using ALP-electron
scattering processes. Recently as well the NEON experiment and the
MINER collaboration have reported upon their searches
\cite{NEON:2024kwv,Mirzakhani:2025bqz}. Although using different
detection technologies compared to those used by the JUNO-TAO
experiment, these results demonstrate the potential of this type of
searches.

In this paper, we claim that upcoming short-baseline neutrino
experiments relying on large organic liquid scintillator
detectors---such as JUNO-TAO \cite{JUNO:2020ijm} or CLOUD
\cite{NavasNicolas:2024ixn}---might contribute to the global ALP
search program. In order to do so, we consider a JUNO-TAO-like
detector concept with parameter specifications as described in
Ref. \cite{JUNO:2020ijm}. For the sake of readability, from here on we
refer to this detector configuration simply as JUNO-TAO. For reasons
to be discussed in Sec. \ref{sec:ALP_prod}, we focus our analysis on
ALP couplings to nuclei and electrons. These couplings induce
scattering and decay processes, with the latter dominating the
``high'' ALP mass region. Thus, their inclusion increases the
discovery potential. The JUNO-TAO detector relies on scintillation
light, which nuclear recoils produce little. We argue that if
scintillation light---along with final-state photons produced in the
scattering process itself---can be recorded over a large enough window
time (one minute or so) an enhancement of the signal might be
possible.

We show as well that both nuclear and electron recoils produce a
prompt and a delayed photon signal, which we argue can potentially be
used for background discrimination. And identify the type of
coincident event signals ALP decays lead to. We demonstrate that
taking advantage of potential background discrimination, JUNO-TAO
might have the capabilities to test regions in the ALP parameter space
where either caveats apply or which have not been yet explored. Our
analysis is, therefore, complementary to the findings provided in
Ref. \cite{Smirnov:2021wgi}.

The remainder of this paper is organized as follows. In
Sec. \ref{sec:ALP_prod}, we provide a discussion of differential
scattering cross sections, decay widths, and survival and decay
probabilities. We present results that allow the calculation of ALP
fluxes and differential event rates. In Sec. \ref{sec:ALP_signals}, we
start with a brief overview of neutrino detection at JUNO-TAO, which
serves as a \textit{frame} for the analysis of scattering ALP signals
in nuclear and electron recoils. We show that ALP signatures imply a
prompt and a delayed photon component if arising from scattering or
coincident photon event signals if generated by ALP decays. In
Sec. \ref{sec:sensitivities}, we present results for sensitivities in
both cases, nuclear and electron recoils. Finally, in
Sec. \ref{sec:conclusions} we summarize and present our
conclusions. Details of a two-dimensional model for the determination
of the most likely ALP decay point within the detector fiducial volume
are presented in {App. \ref{sec:model_ALP_int_FV}.
\section{ALP production at reactors and detection processes}
\label{sec:ALP_prod}
From a phenomenological point of view production of ALPs in reactors
and their further detection depends on whether they couple to photons,
electrons or quarks. Coupling to photons, controlled by the
dimensionful coupling $g_{a\gamma\gamma}$\,, allow production through
Primakoff-like scattering and detection through the corresponding
inverse process. It enables as well ALP decay to photon
pairs. Primakoff-like processes follow from the standard Primakoff
effect, where photon-nucleus scattering produces neutral pions.

Coupling to electrons, in this case controlled by the dimensionless
coupling $g_{aee}$, involve a richer phenomenology. Production goes
through Compton-like processes, while detection through the inverse
process or axio-electric absorption (analogous to photo-electric
absorption). It can generate ALP decays to $e^+e^-$ pairs, in
kinematically allowed regions.

Couplings to quarks imply couplings to nucleons, dictated by the
dimensionless coupling $g_{ann}$. Production can proceed through
different avenues. For instance, heavy isotopes undergoing either
$\alpha$ or $\beta^+$ decay produce excited nuclear isotopes that
through de-excitation will produce monoenergetic ALPs from magnetic
multipole transitions (ALPs are parity odd pseudoscalars,
$J^\text{P}=0^-$). Detection will happen through nuclear resonant
absorption, which will be scarce. The reason being that the flux of
monoenergetic ALPs follow from nuclear de-excitations of a large
variety of nuclear isotopes, of which carbon---the main material
target in the JUNO-TAO detector---contributes only to a small
fraction\footnote{The cumulative fission yields of $^{235}\text{U}$
  and $^{241}\text{Pu}$ include carbon along with other 57 and 54
  nuclear isotopes, respectively \cite{IAEA_livechart}.}.  And
monoenergetic ALPs that do not match the nuclear energy gaps of the
carbon nucleus are very unlikely to get absorbed.

Scenarios where production proceeds through $g_{ann}$ couplings, thus,
typically require an extra coupling for detection. For instance,
pioneer searches at the Bugey reactor in the 80's relied on ALP decays
to photon pairs \cite{Cavaignac:1982ek}. Analysis of these type of
models---therefore---demand going beyond minimality, defined as
scenarios where only one coupling is present at a time. Here we then
focus only on ALP models where the ALP couples either to photons or
electrons.
\subsection{ALP production and detection through
  $g_{a\gamma\gamma}$ couplings}
\label{sec:gagg_couplings}
Photon interaction in the electromagnetic (EM) field of nuclei produce
an ALP flux through the $t$-channel $\gamma+N\to a+N$ process. Its
intensity depends on the photon flux and the ALP-photon conversion
cross section, which for scattering in a nucleus with atomic number
$Z$ reads \cite{Aloni:2019ruo}
\begin{equation}
  \label{eq:ALP_gamma_cross_section}
  \frac{d\sigma^\text{P}_a}{dt}=2\,\alpha\,Z^2\,F^2(t)\,g_{a\gamma\gamma}^2\,
  \frac{m_N^4}{t^2(m_N^2-s)^2(t-4m_N^2)^2}G(t,s)\ .
\end{equation}
Here $\alpha$ is the fine structure constant, $F(t)$} is a form factor
that accounts for electron cloud screening effects, which are relevant
for ALP energies of the order of a few eV, $s$ and $t$ are the
Mandelstam variables, and $m_N$ is the nucleus mass.  Thus, for the
energy range we are interested here ($E_a\gtrsim m_e$, where $E_a$ and
$m_e$ are, respectively, ALP energy and electron mass) ALPs scatter
off the nucleus with no intervention by the electron cloud. Note that
because of the kinematic cut $E_a\lesssim 10\,$MeV, nuclear form
factor effects have no significant impact. The differential cross
section in Eq. (\ref{eq:ALP_gamma_cross_section}), therefore, can be
safely evaluated in the limit $F(t)\to 1$. Furthermore, the kinematic
function $G(t,s)$ reads
\begin{equation}
  \label{eq:kinematic_function_Primakoff}
  G(t,s)=m_a^2\,t\,(m_N^2+s) - m_a^4\,m_N^2 - t\,[(m_N - s)^2 +s\,t]\ .
\end{equation}

The reactor  differential photon flux $d\Phi_\gamma/dE_\gamma$  ($E_\gamma$ being the photon energy)
  involves a continuous
component that follows from electron Bremsstrahlung, Compton
scattering and electron-positron pair annihilation\footnote{Because
  of the relative direction of the initial-state positron and
  final-state photon, the would-be 511 keV spectral line broaden (see
  Sec. \ref{sec:neutrino_detection} for details).}. It involves as
well a host of spectral lines that come from the de-excitation of the
different isotopes fission yields. Since we focus our analysis on ALP
couplings to either photons or electrons, here we consider only the
continuous component which is well described by a power-law spectrum
\cite{bechteler_faissner_yogeshwar_seyfarth_1984}.

ALPs produced in the reactor core propagate isotropically. For a
detector placed at a baseline $L$, ALPs with lifetimes larger than
$L(E_a/|\vec{p}_a|)$ ($\vec{p}_a$ being the ALP momentum) have a
non-zero probability of interacting with the target material. Since
the ALP lifetime is determined by $g_{a\gamma\gamma}$ and $m_a$ (see
below), in some regions of parameter space the flux will be
diminished. This is accounted for by the survival probability
\begin{equation}
  \label{eq:prob_surv}
  \mathcal{P}_\text{Surv} = e^{-L/\ell_a}=e^{-L/(v_a\tau_a)}=e^{-LE_a/(|\vec{p}_a|\tau_a)}\ ,
\end{equation}
where $\tau_a=\Gamma_a^{-1}$ refers to the ALP lifetime and $\Gamma_a$
to its total decay width
\begin{equation}
  \label{eq:total_width_agg}
  \Gamma(a\to \gamma\gamma)=g_{a\gamma\gamma}^2\,\frac{m_a^3}{64\pi}\ .
\end{equation}

The differential cross section in
Eq. (\ref{eq:ALP_gamma_cross_section}) is sharply peaked, as can be
seen by noticing that $|t|/m_N^2=2E_r/m_N\ll 1$ (with $E_r$ the
nucleus recoil energy) and so scattering produces forward scattered
ALPs. This observation allows for the following simplification
\begin{equation}
  \label{eq:Primakoff_cross_section_peaked}
  \frac{d\sigma^\text{P}_a}{dE_a}=\sigma^\text{P}_a(E_a)\delta(E_a-E_\gamma)\ ,
\end{equation}
where the total cross section is obtained from integration of
Eq. (\ref{eq:ALP_gamma_cross_section}). As any flux, the ALP flux
follows from a convolution over photon energies of the differential
cross section in Eq. (\ref{eq:Primakoff_cross_section_peaked}) and the
continuous photon flux. However, the sharp peak structure of the
differential cross section allows for a simplified expression of the
ALP flux \cite{AristizabalSierra:2020rom}
\begin{equation}
  \label{eq:ALP_flux_gagg}
  \frac{d\Phi^\text{P}_a}{dE_a}=\mathcal{P}_\text{Surv}
  \frac{\sigma^\text{P}_a}{\sigma_\text{Tot}}
  \left .\frac{d\Phi_\gamma}{dE_\gamma}\right|_{E_\gamma=E_a}\ .
\end{equation}
Here $\sigma_\text{Tot}=\sigma_\text{SM}+\sigma^\text{P}_a$, with
$\sigma_\text{SM}$ involving all possible EM processes with an
initial-state photon: Compton scattering, Rayleigh scattering,
photo-electric absorption and electron-pair production in the field of
the nucleus and the electron \cite{NIST_XCOM}. It is worth pointing
out that the differential cross section-to-total cross section ratio
``measures'' the amount of photons which undergo ALP conversion. It
can be understood as a branching ratio equivalent. Furthermore, for
$g_{a\gamma\gamma}\lesssim 10^{-3}\,\text{GeV}^{-1}$ (the ALP coupling
parameter space region of interest) the SM cross section exceeds the
Primakoff cross section by several orders of magnitude. The
non-standard contribution can then be safely neglected.

Detection proceeds through inverse Primakoff (IP) scattering
($a+N\to \gamma+N$) and ALP decay (D). The differential event rate can
then be written as
\begin{equation}
  \label{eq:diff_event_rate_Primakoff_decay}
  \left . \frac{dN}{dE_a}\right|_\text{Total} =
  \left . \frac{dN}{dE_a}\right|_\text{IP}
  + \left . \frac{dN}{dE_a}\right|_\text{D}\ .
\end{equation}
Since the inverse Primakoff differential cross section matches that of
Primakoff
scattering---Eq.~(\ref{eq:Primakoff_cross_section_peaked})---up to a
factor 2 \cite{AristizabalSierra:2020rom}, the first term in
Eq. (\ref{eq:diff_event_rate_Primakoff_decay}) can be simplified in
the same way the ALP photon flux has been.  Explicitly both terms
read
\begin{align}
  \label{eq:diff_event_rates_P_scattering_decay}
  \left .\frac{dN}{dE_a}\right|_\text{IP}&=
                                           \frac{m_\text{det}}{2\pi L^2}N_A
                                           \sum_{i=\text{H,C}}
                                           \frac{f_i^n}{m^i_\text{molar}}
                                           \sigma_{a,i}^\text{P}
                                           \left .
                                           \frac{d\Phi_a^\text{P}}{dE_a}
                                           \right|_\text{U}\ ,
  \\
  \left .\frac{dN}{dE_a}\right|_\text{D}&=
                                          \mathcal{P}_\text{Decay}
                                          \frac{A}{4\pi L^2}
                                          \left .
                                          \frac{d\Phi_a^\text{P}}{dE_a}
                                          \right|_\text{U}\,,
\end{align}
where $m_{\text{det}}$ is the detector mass, $N_A$ is the Avogadro
number, $m_{\text{molar}}$ is the target nuclei molar mass and $A$ is
the detector transverse area (see below).  The JUNO-TAO detector is
filled with linear alkylbenzene, LAB in short. Thus, $f_i^n$ refers to
the hydrogen and carbon LAB mass fractions [see
Eq. (\ref{eq:LAB_mass_fractions}) and the discussion thereof as well
as Sec. \ref{sec:neutrino_detection}]. In both cases the ALP flux is
calculated by assuming photon production in uranium. For
$m_\text{det}$ we employ 1 tonne \cite{JUNO:2020ijm}. Note that the
differential event rate in decays involves the probability of decay
within the detector fiducial volume as well as its transverse area
$A=\pi\,R^2_\text{JUNO-TAO}$ ($R_\text{JUNO-TAO} = 0.65$m is the
JUNO-TAO detector fiducial volume radius).  The former is given by
\begin{equation}
  \label{eq:Prob_decay}
  \mathcal{P}_\text{Decay} = 1 - e^{-r E_a/(|\vec{p}_a|\tau_a)} \ ,
\end{equation}
where $r$ refers to the most likely depth where decay might take
place. Note as well that the decay probability increases with
increasing ALP mass, so expectations are that decays become relevant
at higher ALP masses despite the process not having a kinematic
threshold. Of course integration of the total differential event rate
over ALP energies, in the range $[m_e,10\,\text{MeV}]$, determines the
number of events for a given point in parameter space.
\subsection{ALP production and detection through
  $g_{aee}$ couplings}
\label{sec:gaee_couplings}
ALP production in scenarios where the ALP couples to electrons is
determined by the following differential cross section
\cite{Brodsky:1986mi}
\begin{equation}
  \label{eq:Compton_like_x_sec}
  \frac{d\sigma_a^\text{C}}{dE_a}=\frac{Z\,\alpha\,x}{4(s-m_e^2)(1-x)E_\gamma}
  \,g_{aee}^2\,G(x)\ ,
\end{equation}
where the kinematic function $G(x)$ reads
\begin{equation}
  \label{eq:kinematic_function_G}
  G(x)=x-\frac{2m_a^2\,s}{(s-m_e^2)^2}+\frac{2m_a^2}{(s-m_e^2)^2}
  \left(\frac{m_e^2}{1-x} + \frac{m_a^2}{x}\right)\ .
\end{equation}
In the electron rest frame, $s=m_e^2+2E_\gamma\,m_e$ and
\begin{equation}
  \label{eq:x_variable_Compton_like}
  x = 1 - \frac{E_a}{E_\gamma} + \frac{m_a^2}{2E_\gamma\,m_e}\ .
\end{equation}
The ALP flux follows as well from a convolution of the differential
cross section in Eq. (\ref{eq:Compton_like_x_sec}), normalized to
$\sigma_\text{Tot}=\sigma_\text{SM}+\sigma_a^\text{C}$, and the
continuous photon flux. It reads
\begin{equation}
  \label{eq:ALP_flux_Compton}
  \frac{d\Phi_a^\text{C}}{dE_a}=\mathcal{P}_\text{Surv}
  \int^{E_\gamma^\text{max}}_{E_\gamma^\text{min}}
  \frac{1}{\sigma_\text{Tot}}\frac{d\sigma_a^\text{C}}{dE_a}\,
  \frac{d\Phi_\gamma}{dE_\gamma}\,dE_\gamma\ .
\end{equation}
The integration upper limit is determined by kinematic constraints,
$E_\gamma^\text{max}=10\;$~MeV, and
$E_\gamma^\text{min}=(m_am_e-m_a^2/2)/(m_e-m_a)$.  The latter follows
from the $2\to 2$ scattering process kinematics. Here---as in the case
of ALPs coupled to light---$\sigma_\text{SM}\gg \sigma^\text{C}_a$ in
the regions of parameter space of interest. Note that the flux is as
well weighted by the survival probability, but in this case the ALP
decay length is determined by the conditions dictated by the ALP decay
mode $a\to e^+ e^-$ for which the width reads
\begin{equation}
  \label{eq:ALP_toep_em}
  \Gamma(a\to e^+e^-)=\frac{m_a\,g_{aee}^2}{8\pi}
  \sqrt{1 - 4\frac{m_e^2}{m_a^2}}\ .
\end{equation}

As we have pointed out, detection proceeds through inverse
Compton-like (IC) scattering, axio-electric absorption (A) and decay
(D). The total differential event rate thus reads
\begin{equation}
  \label{eq:total_DER_gaee}
  \left . \frac{dN}{dE_a}\right|_\text{Total} =
  \left . \frac{dN}{dE_a}\right|_\text{IC} +
  \left . \frac{dN}{dE_a}\right|_\text{A} +
  \left . \frac{dN}{dE_a}\right|_\text{D}\ .
\end{equation}
For a detector deployed at a distance $L$ from the reactor core, the
different contributions are given by \cite{AristizabalSierra:2020rom}
\begin{widetext}
  \begin{align}
    \label{eq:inverse_Compton_like}
    \left .\frac{dN}{dE_\gamma}\right|_\text{IC}&=\frac{m_\text{det}}{4\pi\,L^2}
                                                  N_A\sum_{i=\text{H,C}}Z_i
                                                  \frac{f_i^n}{m_\text{molar}^i}
                                                  \int_{E_a^\text{min}}^{E_a^\text{max}}
                                                  \left .
                                                  \frac{d\sigma_a^\text{IC}}{dE_\gamma}
                                                  \right|_\text{Z=1}
                                                  \left .
                                                  \frac{d\Phi_a^\text{C}}{dE_a}
                                                  \right|_\text{U}
                                                  dE_a\ ,
    \\[2mm]
    \label{eq:axio-electric}
    \left . \frac{dN}{dE_a}\right|_\text{A}&=\frac{m_\text{det}}{4\pi\,L^2}
                                             N_A\sum_{i=\text{H,C}}
                                             \frac{f_i^n}{m_\text{molar}^i}
                                             \sigma_\text{A}^i
                                             \left .
                                             \frac{d\Phi_a^\text{C}}{dE_a}
                                             \right|_\text{U}\ ,
    \\[2mm]
    \label{eq:decay}
    \left . \frac{dN}{dE_a}\right|_\text{D}&=\frac{A}{4\pi\,L^2}
                                             \mathcal{P}_\text{Decay}
                                             \left .
                                             \frac{d\Phi_a^\text{C}}{dE_a}
                                             \right|_\text{U}\ .
  \end{align}
\end{widetext}
The decay probability in Eq. (\ref{eq:decay}) follows
Eq. (\ref{eq:Prob_decay}), but in this case the decay lifetime is
calculated from the ALP decay mode $a\to e^+e^-$.  Here, we have
written the differential event rate for the inverse Compton-like
process in terms of the final-state photon energy. For the
corresponding expression in terms of the initial-state ALP energy, one
can write
$dN/dE_a|_\text{IC}=|dE_\gamma/dE_a|\,dN/dE_\gamma|_\text{IC}$, with
the Jacobian determined by the constraints implied by the kinematics
of the $2\to 2$ scattering process
\begin{equation}
  \label{eq:Egamma_IC}
  E_\gamma=\frac{y}{2(m_e + E_a - |\vec{p}_a|\cos\theta)}\ ,
\end{equation}
where $y=2m_e E_a + m_a^2$ and $\theta$ is the angle between the incoming
ALP and outgoing photon trajectories. The lower integration limit in
Eq. (\ref{eq:inverse_Compton_like}) follows from
Eq. (\ref{eq:Egamma_IC}) by solving for $E_a$ with $\theta=0$ (and so
$E_\gamma^\text{max}=10\;$MeV). The upper limit, instead, from
Eq. (\ref{eq:x_variable_Compton_like}) by solving for $E_a$ and noting
that $E_a$ is maximized for $E_\gamma=E_\gamma^\text{max}$ and
$x=x_\text{min}$. The latter given by \cite{Chanda:1987ax}
\begin{equation}
  \label{eq:x_min}
  x_\text{min}=\frac{1}{2s}
  \left[
    (s-m_e^2+m_a^2)-\sqrt{(s-m_e^2+m_a^2)^2-4m_a^2s}
  \right]\ .
\end{equation}

The inverse Compton-like differential cross section and the
axio-electric absorption cross section read
\cite{Avignone:1988bv,SuperCDMS:2019jxx}
\begin{widetext}
\begin{align}
  \label{eq:inverse_compton}
  \frac{d\sigma_a^\text{IC}}{dE_\gamma}&=\frac{Z\alpha\,g^2_{aee}}{32\pi}
                                         \frac{y}{m_e^2|\vec{p}_a|^2E_\gamma}
                                         \left(1 +
                                         \frac{4m_e^2E_\gamma^2}{y^2} -
                                         \frac{4m_eE_\gamma}{y} -
                                         \frac{4m_a^2m_e|\vec{p}_a|^2E_\gamma}{y^3}
                                         \sin^2\theta\right)\ ,
  \\
  \label{eq:axio_electric}
  \sigma_\text{A}&=\frac{g_{aee}^2}{|\vec{\beta}|}\frac{3E_a^2}{16\,\pi\,\alpha\,m_e^2}
  \left(1- \frac{|\vec{\beta}|^{2/3}}{3}\right)\sigma_\text{PE}(E_a)\ .
\end{align}
\end{widetext}
Here $\theta$ is obtained from the kinematic constraints implied by
Eq. (\ref{eq:Egamma_IC}), $|\vec{\beta}|$ is the ALP rapidity and
$\sigma_\text{PE}(E_a)$ is the SM photo-electric absorption total
cross section evaluated at ALP energies.

The results presented in this section, along with those from
Sec. \ref{sec:gagg_couplings}, allow the determination of ALP event
rates. There are, however, a couple of questions that still have not
been addressed. First of all, how do signals at the JUNO-TAO detector
look like. Secondly, up to which degree these signals can be
identified and separated from background. The latter comes from
different sources---including neutrino scattering events---but of
which the most abundant are singles from radioactivity that amount to
about $3\times 10^9$ events/year \cite{JUNO:2020ijm}. In what follows,
then, we aim to determine how the signals actually look like and
whether their features enable their identification above
background. We want to stress that the analysis is only aimed at
providing a first approach to the most general problem. A more
realistic and precise determination very likely calls for a full
\texttt{Geant4} detector simulation.
\section{ALP signals at JUNO-TAO}
\label{sec:ALP_signals}
To determine more precisely the type of ALP signals expected at
JUNO-TAO, we rely on the same strategy employed for neutrino
detection. We take advantage of the final-state photon produced in
either inverse Primakoff or inverse Compton-like processes, on which
detection is based upon. Thus, we first review neutrino detection at
JUNO-TAO and then proceed with ALP signals.
\subsection{Electron antineutrino detection at JUNO-TAO}
\label{sec:neutrino_detection}
Electron anti-neutrinos produced at the Taishan Nuclear
Power Plant are detected through inverse $\beta$ decay (IBD)
processes:
\begin{equation}
  \label{eq:IBD}
  \overline{\nu}_e + p^+ \to n^0 + e^+ \ .
\end{equation}
The final-state products produce two distinctive $\gamma$ signals, as
we now discuss \cite{Qian:2018wid}. In the proton rest frame, energy
conservation implies:
\begin{equation}
  \label{eq:energy_conservation_pair_annhilation}
  E_{\overline{\nu}_e} = m_n - m_p + m_e + T_n + T_{e^+}\ ,
\end{equation}
where $T_n$ and $T_{e^+}$ are, respectively, the neutron and positron
kinetic energies. From
Eq. (\ref{eq:energy_conservation_pair_annhilation}) one can see that
at threshold ($T_n=T_{e^+}=0$) IBD will happen provided
$E_{\overline{\nu}_e}\geq
E_{\overline{\nu}_e}^\text{Th}=1.804\;$MeV. For electron-positron pair
annihilation into $\gamma_1+\gamma_2$:
\begin{equation}
  \label{eq:electron_postron_annihilation}
  e^+(p_{e^+})+e^-(p_{e^-})\to\gamma_1(p_\gamma)+\gamma_2(p_\gamma^\prime)\ ,
\end{equation}
energy and momentum conservation (in the electron reference frame)
implies
\begin{align}
  \label{eq:e+_e-_to_g_+_g}
  T_{e^+}+2m_e=E_\gamma^\prime+E_\gamma\ ,
  \\
  \label{eq:momentum_cons}
  \vec{p}_{e^+}=\vec{p}_\gamma+\vec{p}^\prime_\gamma\ .
\end{align}
At threshold, this means that pair annihilation produces back-to-back
monochromatic photons with $E_\gamma=E_\gamma^\prime=m_e=511$~keV,
i.e. monochromatic lines at $\lambda=2.4\,\text{pm}$. For neutrinos
above threshold ($T_{e^+}\neq 0$), photon energies become angular
dependent
\begin{align}
  \label{eq:Egamma_IBD}
  E_\gamma &= \frac{m_e(T_{e^+}+2m_e)}{T_{e^+}+2m_e-|\vec{p}_{e^+}|\cos\theta}\ ,
  \\[2mm]
  \label{eq:Egamma_prime_IBD}
  E_\gamma^\prime&=\frac{(T_{e^+}+2m_e)(T_{e^+}+m_e-|\vec{p}_{e^+}|\cos\theta)}
                   {T_{e^+}+2m_e-|\vec{p}_{e^+}|\cos\theta}\ .
\end{align}
Here $\theta$ refers to the angle between the incoming positron and
 one of the outgoing photons ($\gamma_1$). From Eqs. (\ref{eq:Egamma_IBD}) and
(\ref{eq:Egamma_prime_IBD}) one can see that when $\gamma_1$ scatters
along the incoming positron direction [forward scattering
($\theta=0$)] $E_\gamma=E_\gamma^\text{Max}$ and so
$E_\gamma^\prime=E_\gamma^{\prime\text{Min}}$. Conversely, when
$\gamma_1$ scatters opposite to the incoming positron direction
[backward scattering ($\theta=\pi$)] $E_\gamma=E_\gamma^\text{Min}$
and so $E_\gamma^\prime=E_\gamma^{\prime\text{Max}}$. Thus, the
angular dependence leads to photon energy distributions that range
from $E_\gamma^\text{Min}$ to $E_\gamma^\text{Max}$.

\begin{figure*}
  \centering
  \includegraphics[scale=0.5]{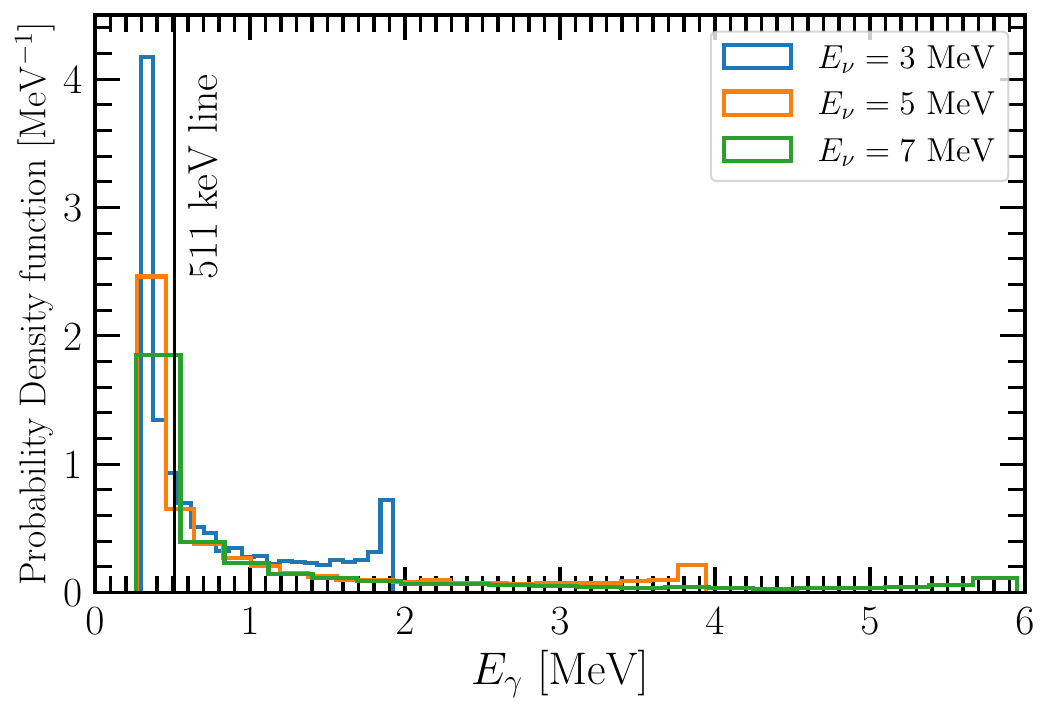}
  \includegraphics[scale=0.5]{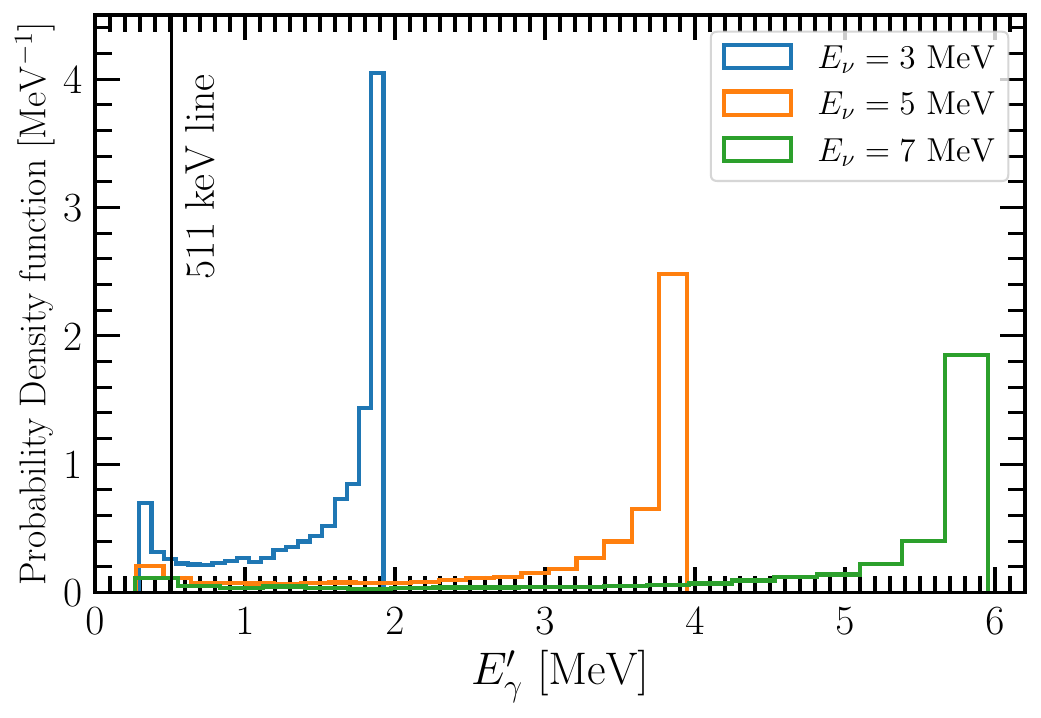}
  \caption{Photon energy probability distribution functions (PDFs) in
    pair annihilation for representative electron antineutrino
    energies (all of them substantially above threshold) for
    $\gamma_1$ (left graph) and $\gamma_2$ (right graph). These
    results are based on random sampling of the kinematic relations
    shown in Eqs. (\ref{eq:Egamma_IBD}) and
    (\ref{eq:Egamma_prime_IBD}).  For the PDFs $10^4$ scattering
    events and 20 energy bins have been assumed. The latter assuming a
    $\Delta E_\gamma\simeq 8\,$keV photon energy
    resolution. Back-to-back photons cluster towards low and high
    energy bins.}
  \label{fig:electron_positron_annihilation_gamma_spectrum}
\end{figure*}
Assuming different incoming electron anti-neutrino energies and $10^4$
scattering events, we have randomly sampled Eqs. (\ref{eq:Egamma_IBD})
and (\ref{eq:Egamma_prime_IBD}) over $\theta\subset [0,\pi]$. With the
results of such sampling, we have then determined the photon energy
probability density functions (PDFs) for both $\gamma_1$ and
$\gamma_2$.
Results are displayed in
Fig.~\ref{fig:electron_positron_annihilation_gamma_spectrum}. For
$\gamma_1$ (left graph) one can see that most events tend to cluster
towards small energy bins. This means that most of those photons are
produced opposite to the incoming positron direction. In
contrast---and as expected---the other final-state photons
($\gamma_2$) clump towards the highest energy bins, in agreement with
our previous discussion (see the right graph of Fig.~\ref{fig:electron_positron_annihilation_gamma_spectrum}).
Note that these do not reflect the event rate angular PDFs, whose
  calculation requires the electron-positron annihilation differential
  cross section. They are just photon energy PDFs derived from simple
  kinematic criteria, to give an approximate qualitative illustration.
Despite these scattering events taking place
above threshold, final-state photons are still back-to-back. Momentum
conservation in Eq. (\ref{eq:momentum_cons}) implies
\begin{equation}
  \label{eq:alpha_theta_pair_annihilation}
  \sin\alpha=\frac{E_\gamma}{E_\gamma^\prime}\sin\theta\ ,
\end{equation}
where $\alpha$ measures the direction of the incoming positron and
$\gamma_2$.  Thus, taking the case of those events for which
$E_\gamma=E_\gamma^\text{Min}$,
Eq. (\ref{eq:alpha_theta_pair_annihilation}) shows that $\alpha=0$.

The photon signal from the final-state neutron proceeds in a different
way. The JUNO-TAO detector uses LAB, a liquid scintillator organic
compound with chemical composition
$\text{C}_6\text{H}_5\text{C}_n\text{H}_{2n+1}$ (with
$n\subset [10,16]$) \cite{JUNO:2020ijm,Steiger:2022fhu}. LAB-based
scintillators have been extensively used in neutrino detection in the
last 15 years or so, Daya Bay along with RENO and COSINE-100 are clear
examples \cite{DayaBay:2012aa,RENO:2010vlj,Adhikari:2017esn}. JUNO-TAO
and JUNO as well thus leverage on this expertise. Although the bulk of
the liquid scintillator is LAB, it comprises as well small fractions
of other compounds: (i) A gadolinium mass fraction of 0.1\%, (ii) a
fluorescent agent (2.5-Diphenyloxazole, PPO,
$\text{C}_{15}\text{H}_{11}\text{N}\text{O}$) concentration of 2 g/L,
(iii) a wavelength shifter (Bis-MSB, $\text{C}_{24}\text{H}_{22}$)
concentration of 1 mg/L, (iv) a co-solvent of 0.05\% ethanol
($\text{C}_2\text{H}_6\text{O}$) \cite{JUNO:2020ijm}.

Of particular relevance for electron anti-neutrino detection is the Gd
doping. Gd has seven stable isotopes, of which five have relative
abundances in the $15\%-25\%$ range. Furthermore it has the largest
neutron absorption cross section among natural stable isotopes, with
$^{155}$Gd and $^{157}$Gd exhibiting the largest cross sections
\cite{NNDC}:
\begin{align}
  \label{eq:Gd_cross_section}
  ^{155}\text{Gd}(n,\gamma):\quad &
                                    \sigma=60740.2\,\,\text{barn}\ ,
                                    \nonumber\\[2mm]
  ^{157}\text{Gd}(n,\gamma):\quad &
                                    \sigma=253765.6\,\,\text{barn}\ .
\end{align}
Thus, neutrons produced in IBD are captured predominantly by
$^{155}\text{Gd}$ and $^{157}\text{Gd}$. Neutron capture produces
either $^{156}\text{Gd}$ or $^{158}\text{Gd}$ excited isotopes which
then cascade decay to their ground state through the emission of
multiple photons. The de-excitation process is a complex one, however
the overall energy of the emitted photons is fixed by the $Q$ value of
the nuclear reaction (neutron binding energy):
\begin{align}
  \label{eq:neutron_capture_photon_emission}
  ^{155}\text{Gd} + n^0
  \to ^{155}\text{Gd}^*&\to ^{156}\text{Gd}
  + \sum_i \gamma_i\;(Q=8.536\,\text{MeV})\ ,
  \nonumber\\
  ^{157}\text{Gd} + n^0\to ^{157}\text{Gd}^*&
  \to ^{158}\text{Gd}
  + \sum_i \gamma_i\;(Q=7.937\,\text{MeV})\ .
\end{align}
Photons produced in the cascade thus produce a second scintillation
light signal that enables the identification of the IBD signal above
background. For a 0.1\% Gd-doped LAB, the average neutron capture
time is $\sim 30\,\mu$s \cite{Qian:2018wid,JUNO:2020ijm}. Thus, the
prompt scintillation signal from pair annihilation combined with the
delayed scintillation signal from neutron capture allows for an
efficient reconstruction of electron antineutrino events.

\begin{figure}
  \centering
  \includegraphics[scale=0.47]{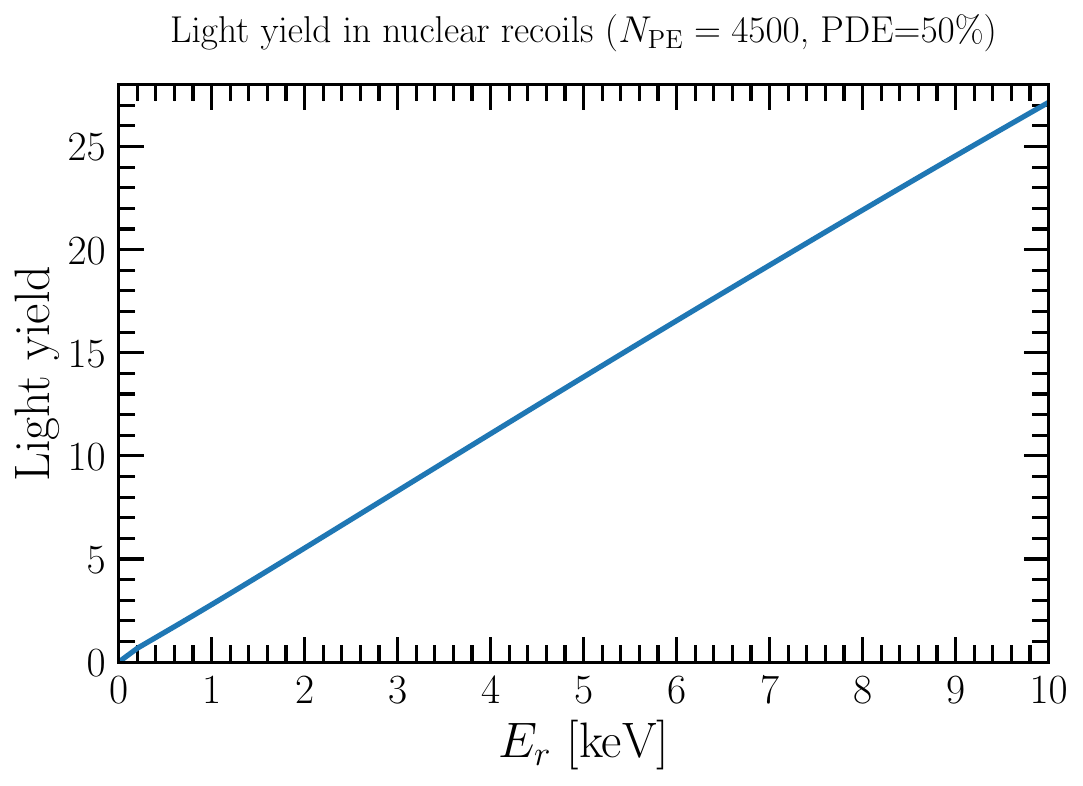}
  \caption{Estimated light yield---produced in a single nuclear
    recoil---as a function of nuclear recoil energy. We use 4500
    photoelectrons per MeV (9 photons per keV) as a proxy and a PDE of
    50\% \cite{JUNO:2020ijm}.}
  \label{fig:light_yield_nuclear_recoil}
\end{figure}
\subsection{ALP detection at JUNO-TAO}
\label{sec:alp_detection}
Leveraging on the same strategy used for neutrino detection, here we
aim at showing that ALP detection features as well prompt and delayed
photon signals. That this is the case is somewhat expected. Both 
inverse Primakoff and inverse Compton-like scattering produce a
final-state photon. So, interactions produce a prompt scintillation
signal from the final-state photon and a delayed one from ionization
produced by nuclear or electron recoil. In what follows, we
characterize both signals and determine their time delay.

To estimate the scintillation light produced in nuclear or electron
recoils, two quantities are required: (i) A representative photon
yield,  (ii) a typical photon emission spectrum.  The photoelectron
yield is about 4500 per MeV. Since photodetection efficiency (PDE)
peaks at 50\%, this means that about 9000 photons per MeV are produced~\cite{JUNO:2020ijm}. The emission spectrum is dominated by the
wavelength shifter BisMSB and peaks at 430 nm, right where the PDE
peaks as well (see e.g.~\cite{Steiger:2022fhu}).
\subsubsection{Inverse Primakoff scattering and ALP decay to photons}
\label{sec:inv_primakoff}
Starting with \textit{Primakoff scattering}---and so nuclear
recoils---the light yield can be calculated with the aid of Birks' law
\cite{Birks:1951boa}
\begin{equation}
  \label{eq:birks_law}
  L_{\text{yield}}(E_r) = \frac{E_r}{1 + k_B^\text{Eff}E_r}Y \ .
\end{equation}

Here $Y$ refers to a proxy light yield for which we use
$Y=9\;\text{photons}/\text{keV}$ and $k_B^\text{Eff}=k_B/\ell$ is an
effective Birks' constant (Birks' constant per particle path length,
$\ell$). Note that in reality Eq. (\ref{eq:birks_law}) follows from
Birks' law after integration over particle path, assuming that energy
loss per unit length is proportional to energy. For Birks' constant we
have
$\rho\,k_B\subset [1,5]\times
10^{-3}\,\text{g}/\text{cm}^2/\text{MeV}$ \cite{Grupen:2008zz}, so
taking $\rho_\text{LAB}=863\times 10^{-3}\,\text{g}/\text{cm}^3$ we
find
\begin{equation}
  \label{eq:k_B_LAB}
  k_B\subset [1.16,5.79]\times 10^{-3}\,\frac{\text{cm}}{\text{MeV}}\ .
\end{equation}
Calculation of $k_B^\text{Eff}$ requires a particle path length (a
function of particle energy). We have adopted the particle length of alpha particles
in plastic scintillator (vinyltoluene-based) \cite{NIST}, which, to a
certain degree, resembles
LAB. Fig. \ref{fig:light_yield_nuclear_recoil} shows the estimated
light yield per nuclear recoil as a function of nuclear recoil energy.

As expected, large light yields demand high recoil energies, which are
by default small for the ALP energies involved: From the kinematics of
the inverse Primakoff process, one finds
\begin{equation}
  \label{eq:Egamma_inverse_Primakoff}
  E_\gamma = \frac{m_a^2+2E_am_N}{2\left[(E_a+m_N)-|\vec{p}_a|\cos\theta\right]}\ ,
\end{equation}
where in this case $\theta$ refers to the angle between the trajectory
of the incoming ALP and the trajectory of the outgoing photon. From
this expression---in the limit $m_a<E_a \ll m_N$---one finds
\begin{equation}
  \label{eq:nuclear_recoil_energy}
  E_r = E_a - E_\gamma \simeq \frac{E_a^2}{m_N}
  \left(1 - \cos\theta\right)\subset \frac{E_a^2}{m_N}[0,2]\ ,
\end{equation}
which shows that $E_r\simeq \mathcal{O}(\text{keV})$. One can see that
maximization of $E_r$ requires backward photon scattering, which is
unlikely because of the initial-state momentum and the mass of the
scatterers in LAB (mainly carbon nuclei). Indeed it is the other way
around, as one can check from the constraints implied by momentum
conservation. In the limit $E_r\ll E_a$, the angle between the
incoming ALP and the outgoing photon can be written as
\begin{equation}
  \label{eq:costheta}
  \cos\theta \simeq 1 - \frac{m_NE_r}{E_a^2} + \frac{E_r}{E_a}\ .
\end{equation}
This demonstrates that inverse Primakoff processes produce
predominantly forward photon scattering, as we have already argued
based on the differential cross section in
Eq. (\ref{eq:ALP_gamma_cross_section}). Recoil energies, therefore,
are expected to cluster towards the lower side of the energy interval
in Eq. (\ref{eq:nuclear_recoil_energy}).

\begin{figure}[h]
  \centering
  \includegraphics[scale=0.47]{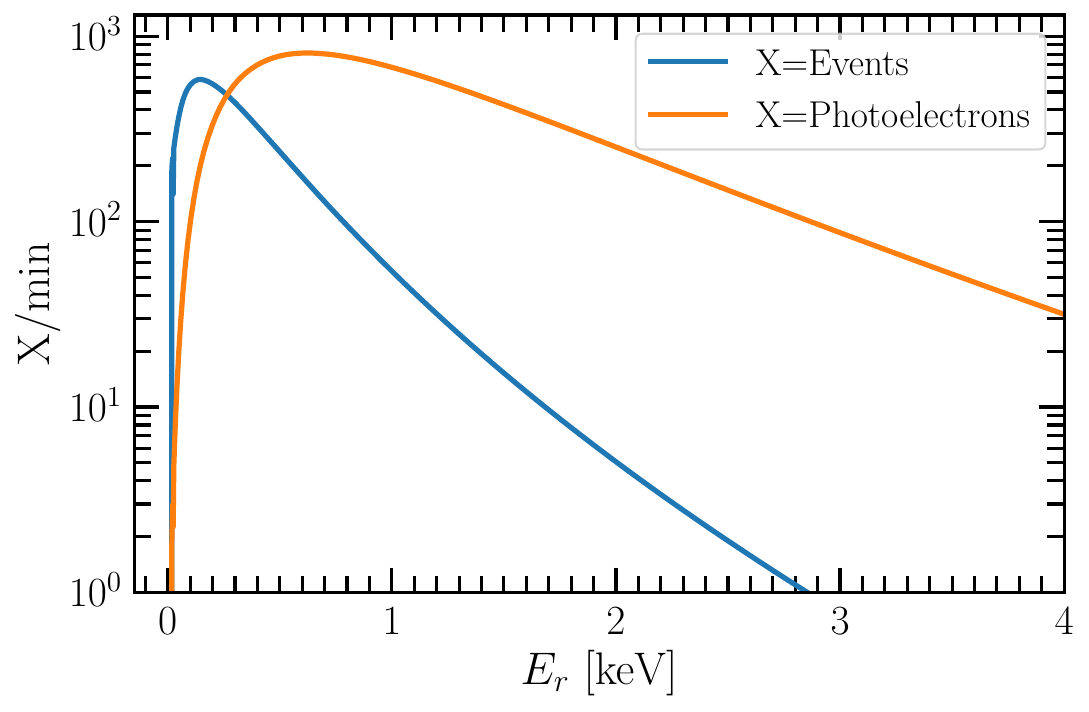}
  \caption{Event rate per minute as well as number of PEs in detection
    through inverse Primakoff scattering as a function of nuclear
    recoil energy.  A point in parameter space within the cosmological
    triangle has been used (see text for more details):
    $m_a=0.3\,\text{MeV}$ and
    $g_{a\gamma\gamma}=9.5\times 10^{-3}\,\text{GeV}^{-1}$.}
  \label{fig:event_rate_Primakoff}
\end{figure}
In principle, however, this does not mean that the photon yield
produced in nuclear recoils is not observable. The point here is that
within a given time interval, say $\Delta t$, a certain amount of
events at fixed recoil energies will take place. So, despite the small
light yield produced in a single nuclear recoil event the overall
light yield over $\Delta t$ may be potentially large enough for the
signal to be recorded. Fig. \ref{fig:event_rate_Primakoff} shows the
expected event rate per minute for a point in parameter space within
the cosmological triangle (see Sec. \ref{sec:sensitivities} for
details) corresponding to $m_a=0.3\,\text{MeV}$ and
$g_{a\gamma\gamma}=9.5\times 10^{-3}\,\text{GeV}^{-1}$. This result
shows that up to $\sim 2\,$keV events are abundant, thus enhancing the
intensity of the scintillation light signal and so of the PE rate.

With the number of events at hand, the number of photons produced in
$\Delta t$ can be calculated according to
$N_\gamma/\Delta t=(\text{Events}/\Delta t)\times L_{\text yield}$. The intensity of
the signal in $\Delta t$ can then be cast as follows
\begin{equation}
  \label{eq:intensity}
  I=\left(\frac{N_\gamma\times L_{\text{yield}}}{10}\right)\;
  \left(\frac{\text{min}}{\Delta t}\right)
  \left(\frac{E_r}{\text{MeV}}\right)
  \left(\frac{10\;\text{m}^2}{A}\right)\;
  \frac{\text{MeV}}{\text{m}^2\;\text{min}}\ .
\end{equation}
Note that here we have taken 10~$\,\text{m}^2$ as the total area
covered by the silicon photo-multipliers \cite{Xu:2022mdi}. The
intensity combined with the PE-per-MeV proxy
($N_\text{PE} = 4500\;\text{PE/MeV}$) allows the determination of the
photoelectron rate,
$N_\text{PE}^{\Delta t}=I\times A\times
N_\text{PE}$. Fig. \ref{fig:event_rate_Primakoff} shows the result as
a function of nuclear recoil energy. It demonstrates that if the
number of photoelectrons over---say---1 min can be
recorded/identified, nuclear recoil events might be observable in
certain regions of parameter space. For this to be the case, however,
the scintillation light spectrum should be characterized.

\begin{figure}
  \centering
  \includegraphics[scale=0.47]{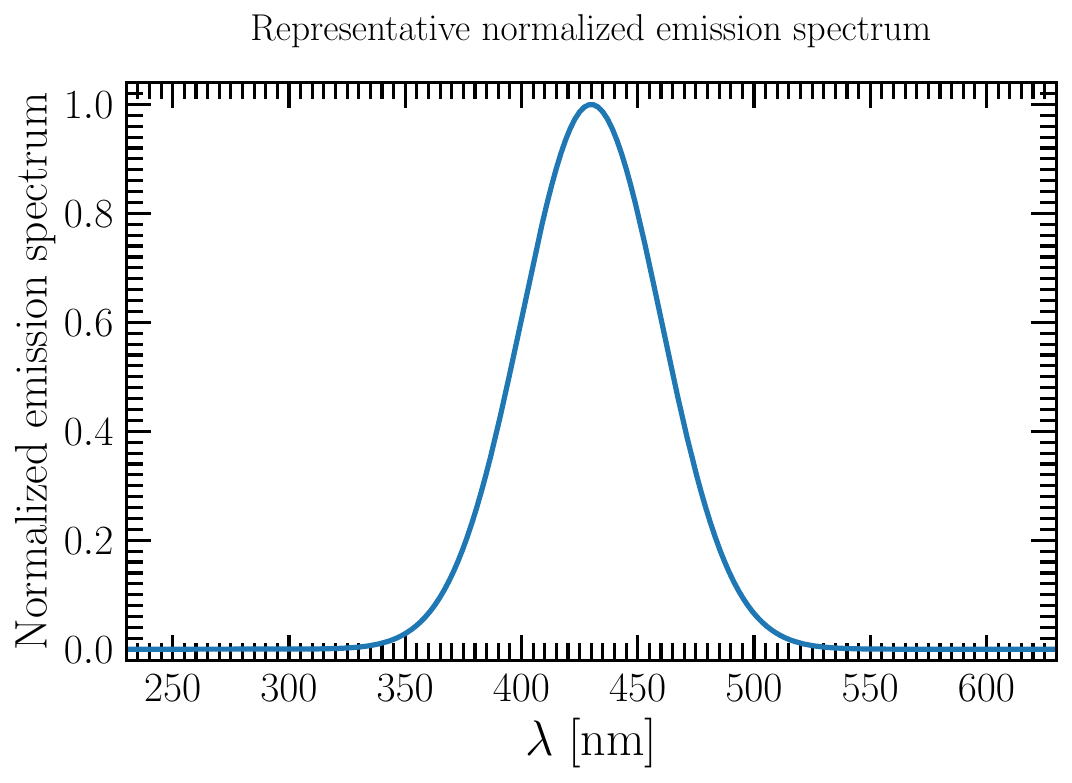}
  \caption{Normalized Gaussian function (by setting $I_0$ = 1 in Eq.~\ref{eq:normalized_intensity})
    with $\mu=430\,$nm as a
    representative photon emission spectrum for the JUNO-TAO
    cocktail.}
  \label{fig:emission_spectrum}
\end{figure}

As we have already stressed, a full characterization of the nuclear
recoil scintillation light spectrum requires an emission spectrum
proxy.  A characterization of the BisMSB emission spectrum shows two
peaks, with the dominant peak at $\lambda=430\,$nm (see
e.g. \cite{Steiger:2022fhu}). Although available, we have instead
restored to a Gaussian function parametrization with mean
$\lambda_0=430\,$nm and standard deviation
$\sigma=30\,\text{nm}$. Thus, for the emission spectrum intensity
(normalized to an arbitrary reference intensity $I_0$ at $\lambda = \lambda_0$) we write
\begin{equation}
  \label{eq:normalized_intensity}
   I(\lambda) = I_0\,
  e^{-\left(\frac{\lambda - \lambda_0}{\sqrt{2}\;\sigma}
    \right)^2}\ .
\end{equation} 
Fig. \ref{fig:emission_spectrum} shows the result we have relied upon
in the determination of the nuclear recoil scintillation light
emission spectrum PDF. Note that although used for practical purposes,
it resembles closely the measured spectrum.

We proceed then as follows. A recoiling nucleus produces excitation of
the surrounding medium, which, through de-excitation processes produce
scintillation light (photons). The amount of photons is determined by
Eq. (\ref{eq:birks_law}) and their wavelengths by the range dictated
by the emission spectrum in Fig. \ref{fig:emission_spectrum},
$[350,500]\,\text{nm}$. We then determine the statistical weight of
the wavelengths within that range by generating a set containing
$N_\lambda=5000$ values and calculating their discrete probability
distribution
\begin{equation}
  \label{eq:discrete_prob_dist}
  \mathcal{P}_i=\frac{{I}(\lambda_i)}
  {\sum_{j=1}^{N_\lambda} {I}(\lambda_j)}\ .
\end{equation}

\begin{figure*}
  \centering
  \includegraphics[scale=0.465]{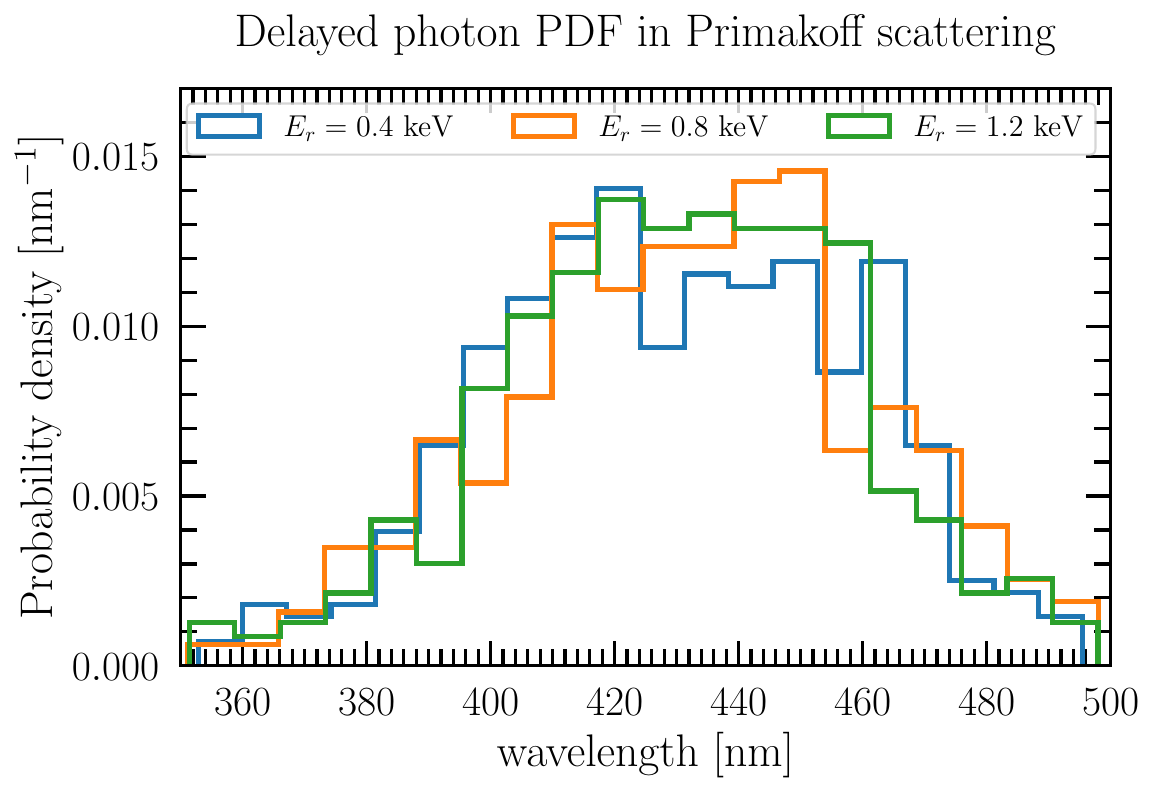}
  \includegraphics[scale=0.465]{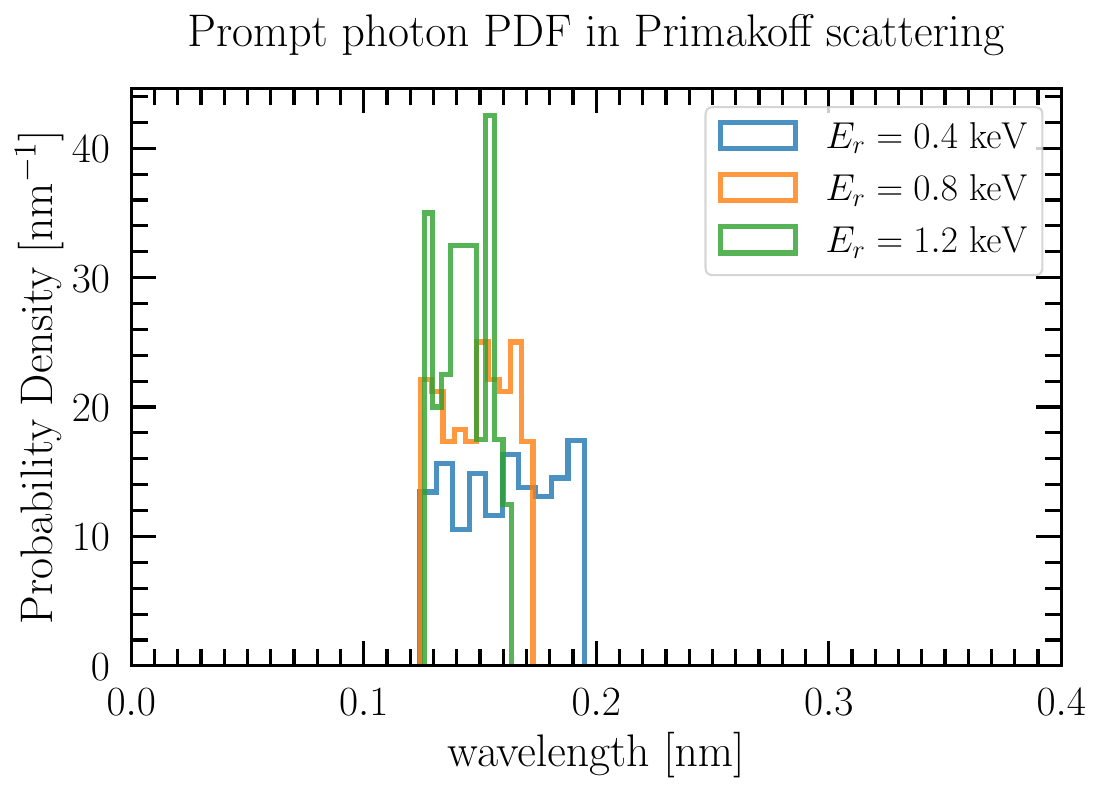}
  \caption{\textbf{Left graph}: Probability distribution function as a
    function of the delayed photon wavelength for nuclear recoil
    scintillation light emission spectrum at different recoil energies
    (delayed photon signal). \textbf{Right graph}: Photon kinematic
    spectrum probability distribution function as a function of the
    prompt photon wavelength for light emission in inverse Primakoff
    processes $a + N\to \gamma + N$ (prompt photon signal).}
  \label{fig:PDF_delayed_prompt_Primakoff}
\end{figure*}

For fixed recoil energies, we then randomly sample from the
distribution in Eq. (\ref{eq:discrete_prob_dist}) as many wavelengths
as photons produced in $\Delta
t$. Fig.~\ref{fig:PDF_delayed_prompt_Primakoff} (left graph) shows the
result for three recoil energy choices (see
Tab. \ref{tab:recoil_energies_num_gammas}), for the same parameter
space point used in the generation of the event spectrum in
Fig. \ref{fig:event_rate_Primakoff}. As expected, the scintillation
light spectrum produced in nuclear recoils follows closely the
emission spectrum proxy (peaking around 430 nm). Thus, spectra of this
type with a relatively small number of photoelectrons recorded over a
sizable time interval---for instance 1 min---could be hinting towards
ALP signals induced by Primakoff scattering. Of course, radioactive
processes are very likely capable of inducing this type of signals as
well (see e.g. Ref. \cite{Xu:2022mdi}). So the question is whether
they can be disentangled from the actual signals.
\begin{table}[h]
  \centering
  \renewcommand{\arraystretch}{1.3}
  \renewcommand{\tabcolsep}{0.5cm}
  \begin{tabular}{|c||c|c|c|}\hline
    \multicolumn{4}{|c|}{Events and $N_\gamma$ at different
    recoil energies}\\\hline\hline
    $E_r$ [keV] & 0.4 & 0.8 & 1.2\\\hline
    Events/min & 325 & 95 & 32 \\\hline
    $N_\gamma/\text{min}$ & 389 & 215 & 106\\\hline
  \end{tabular}
  \caption{Number of events and photons generated in nuclear recoils
    at different recoil energies for $\Delta t = 1\,$min and for the
    parameter space point $m_a=0.3\,$MeV and
    $g_{a\gamma\gamma}=9.5\times 10^{-5}\,\text{GeV}^{-1}$. These
    values are used to generate the scintillation light spectra shown
    in Fig. \ref{fig:PDF_delayed_prompt_Primakoff}.}
  \label{tab:recoil_energies_num_gammas}
\end{table}

As we have already anticipated, the fact that the scattering process
itself produces final-state photons might enable their identification
(at least partially). In contrast to scintillation light produced in
nuclear recoils---that requires a certain time to build up---these
photons are produced directly in the scattering process. This implies
a time delay that could be used to separate signal from background,
in the same way it is done in electron anti-neutrino detection. In other
words, here we want to argue that final-state photons produce a prompt
photon component while nuclear recoils a delayed one.

The identification of the prompt signal requires at least the
characterization of the photon kinematic spectrum PDF. This can be
done by noting that final-state photons within a narrow energy range
produce nuclear recoils with fixed recoil energy. Since
$E_a\simeq E_\gamma$, that range can be determined with the aid of
Eq.~(\ref{eq:nuclear_recoil_energy})
\begin{equation}
  \label{eq:Eg_min_Eg_max}
  E_\gamma=\sqrt{\frac{m_NE_r}{1-\cos\theta}}
  \subset \sqrt{\frac{m_NE_r}{2}}
  \left[
    \frac{1}{\sin(\theta_\text{max}/2)},
    \frac{1}{\sin(\theta_\text{min}/2)}
  \right]\ .
\end{equation}
Here $\theta_\text{min}$ is fixed by the kinematic constraint
$E_\gamma^\text{max}=10\,$MeV. So, given a certain recoil energy one
finds
\begin{equation}
  \label{eq:theta_min}
  \theta_\text{min}=2\arcsin\left(\frac{1}{E_\gamma^\text{max}}
  \sqrt{\frac{m_N E_r}{2}}\right)\ .
\end{equation}
To determine $\theta_\text{max}$ we adopt
$\theta_\text{max} = \theta_\text{min}+\Delta\theta$, with
$\Delta \theta=\pi/18$. This choice is motivated by the observation that
Primakoff scattering produces mostly forward photons. With the photon
energy range fixed, and for the recoil energy values and corresponding
events in Tab. \ref{tab:recoil_energies_num_gammas}, we have randomly
sampled $E_\gamma$.

Results for the prompt photon kinematic spectrum PDF are shown in
Fig.~\ref{fig:PDF_delayed_prompt_Primakoff} (right graph). In contrast
to the scintillation light emission PDF, whose profile follows the
Gaussian emission profile, in this case the spectra are rather flat,
narrow, and in the X-ray part of the electromagnetic spectrum. Note
that since the JUNO-TAO cocktail involves a wavelength shifter
(BisMSB), these wavelengths might shift towards larger values.

\begin{figure}[h]
  \centering
  \includegraphics[scale=0.4]{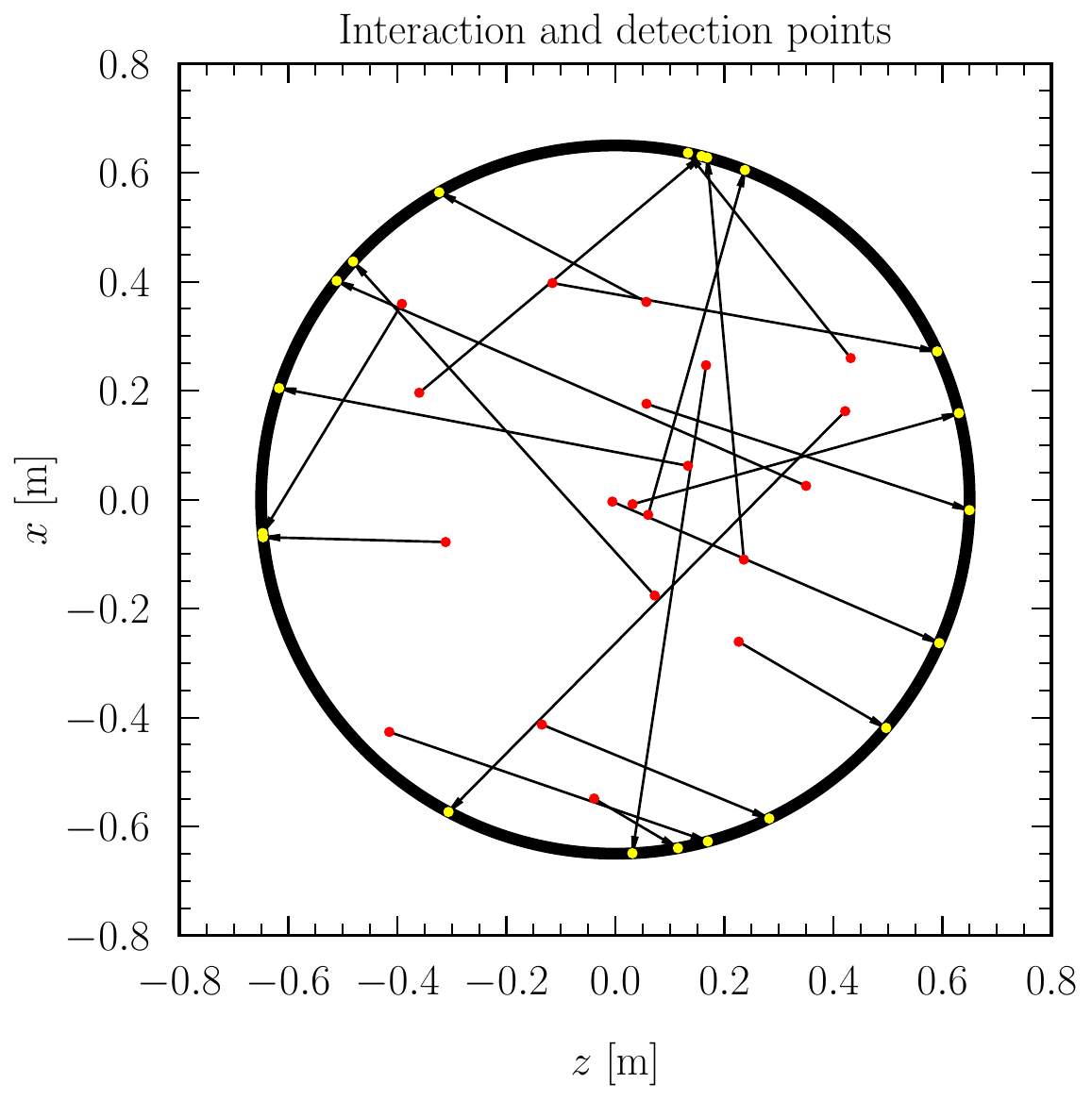}
  \caption{Schematic sample of final-state photons production points
    (red) and detection points (yellow). This graph illustrates the
    procedure we have employed in the determination of the survival
    probability for a given fraction of photons.}
  \label{fig:schematic_detector}
\end{figure}
Having characterized the prompt photon kinematic spectrum PDF, we are
now in a position to quantify the time delay between the two
signals. In general, photons produced in nuclear (or electron) recoils
and/or final-state photons produced in the ALP-nucleus (or electron)
interactions have a non-zero likelihood of being reprocessed by the
medium (LAB). There is also a non-zero probability that some of those
photons reach a SiPM with no further interaction. These fractions can
be estimated with the aid of a simplified version of production and
detection, as we now discuss.

To do so, we have randomly sampled over a two-dimensional projection of
the JUNO-TAO fiducial volume. This allows the generation of production
points. We then randomly sample detection points over the active
volume boundary. Production and detection points determine the photon
trajectory lengths, which can then be compared with the photon mean
free path of the compound (assumed to be
$\text{C}_6\text{H}_5\text{C}_{13}\text{H}_{27}$, for definiteness),
extracted from Ref. \cite{NIST_XCOM}. A schematic sample of the output
is shown in Fig. \ref{fig:schematic_detector}.

\begin{figure*}[t]
  \centering
  \includegraphics[scale=0.47]{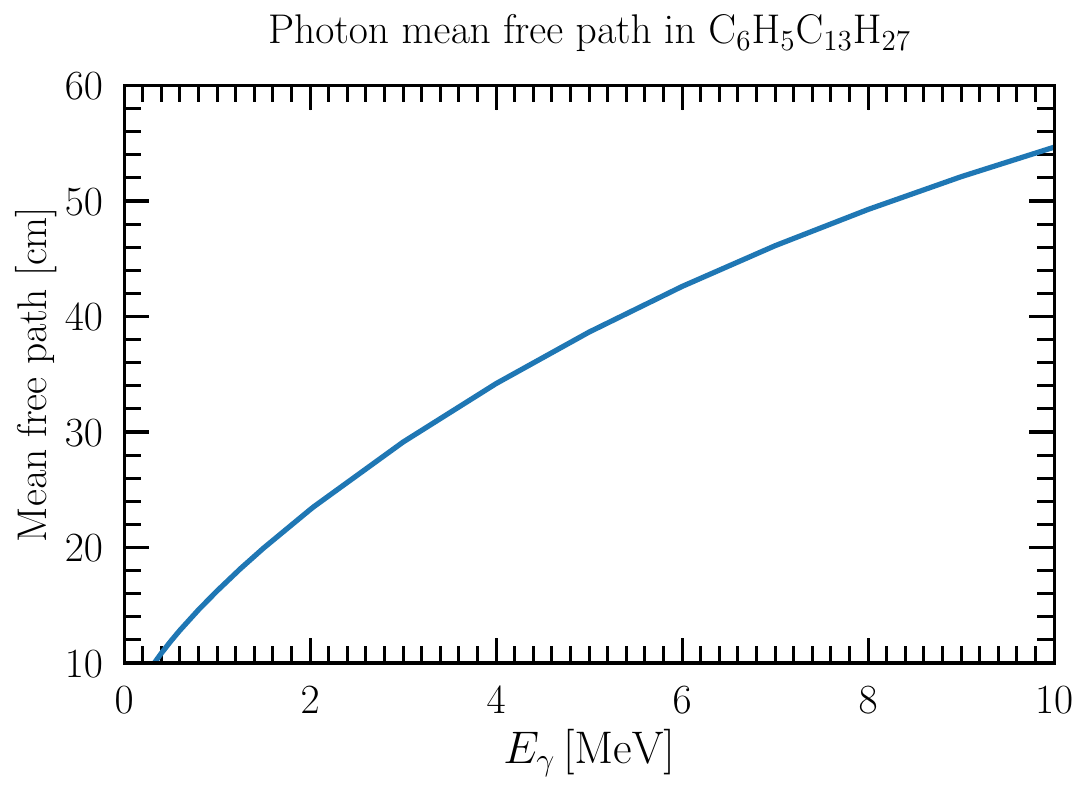}
  \includegraphics[scale=0.47]{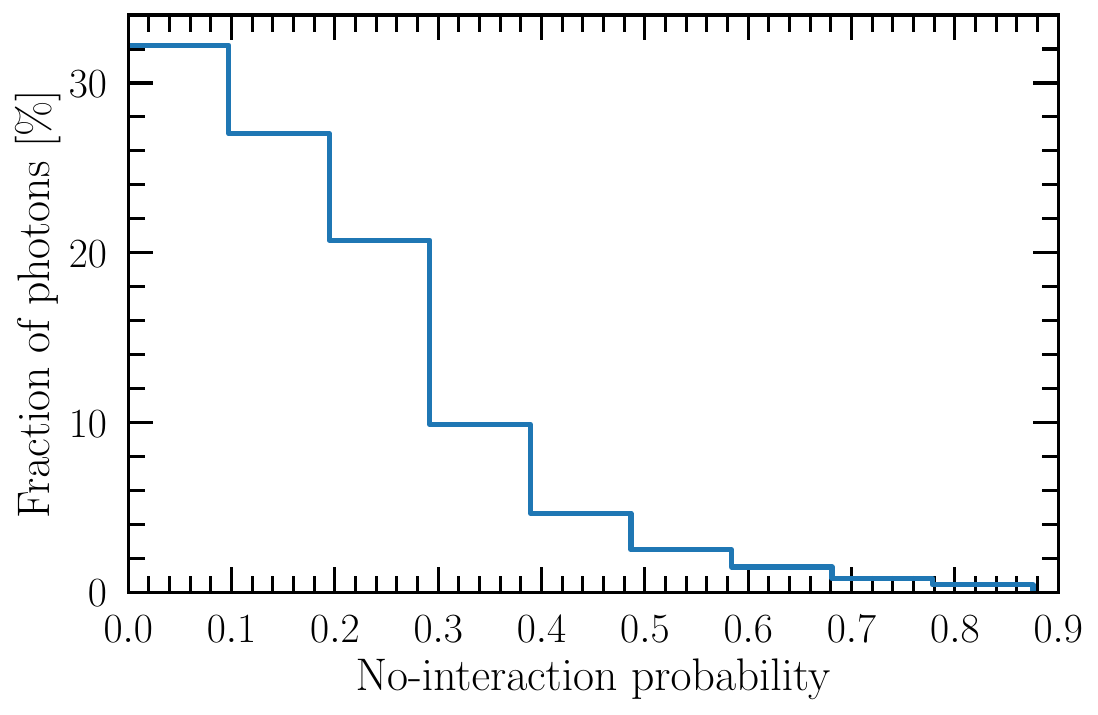}
  \caption{\textbf{Left graph}: Photon mean free path in
    $\text{C}_6\text{H}_5\text{C}_{13}\text{H}_{27}$ as a function of
    photon energy \cite{NIST_XCOM}. \textbf{Right graph}: Fraction of
    photons that reach a SiPM without further interacting with the
    medium. From this result one can see that $\sim 5\%$ of those
    photons have a survival probability in the range $40\%-50\%$.}
  \label{fig:mean_free_path_and_fraction_photons}
\end{figure*}
Randomly assigning energies to those events, the survival probability
for photon propagation can be written as
\begin{equation}
  \label{eq:survival_Prob}
  \mathcal{P}(E_\gamma)=e^{-d(E_\gamma)/\ell(E_\gamma)}\ ,
\end{equation}
where $d(E_\gamma)$ stands for the length of each photon trajectory
and $\ell(E_\gamma)$ to the photon mean free path in
$\text{C}_6\text{H}_5\text{C}_{13}\text{H}_{27}$. Sampling over
Eq. (\ref{eq:survival_Prob}) ---after constructing the sample
$d(E_\gamma)$--- allows the determination of the fraction of photons
that have a certain probability of reaching a SiPM without further
interacting with the target material. Results are displayed in
Fig. \ref{fig:mean_free_path_and_fraction_photons}. The left graph
shows the photon mean free path, while the right graph shows the
fraction of photons in terms of their survival probability. From the
latter, one can see that about $5\%$ of those photons have a 40\%-50\%
probability of reaching a SiPM without further interacting with the
medium.

There is, of course, the question of what are the most probable photon
energies for which that will be the case. Given the expression for the
survival probability---Eq. (\ref{eq:survival_Prob})---and the photon
mean free path, it is clear that the most likely photons to get from
production to detection without interacting with the medium are close
to the high energy tail. Those photons are produced by the most
energetic ALPs, which in turn are less abundant. So, despite having a
relatively large fraction, they are expected to be less abundant just
because of kinematic criteria.

A photon produced in an interaction taking place in a random point
$d\subset [0,0.65]\,$m within the active volume will reach a SiPM in
a time given by $t_\text{int} = d/v_\gamma$, with
$v_\gamma=c/n_\text{LAB}$ and $n_\text{LAB}$ the refractive index for
LAB \cite{Yeo:2010zz}. A recoiling nucleus---instead---will transfer
energy to the medium during $\tau_\text{recoil}$, after which the
medium will undergo de-excitation during $\tau_\text{decay}$. The
scintillation photons produced in these processes will then reach the
SiPM after $t_\text{int}$. One can then write
\begin{align}
  \label{prompt_and_delayed_signals}
  t_\text{prompt} & = t_\text{int}\ ,
                    \nonumber\\
  t_\text{delayed} & = \tau_\text{recoil} + \tau_\text{decay} + t_\text{int}\ .
\end{align}
The time elapsed between the prompt and the delayed signals can then
be estimated to be
$\tau=t_\text{delayed}-t_\text{prompt}=\tau_\text{recoil} +
\tau_\text{decay}$.  Here we have assumed that the photons are not
reprocessed by the medium (a small fraction will, as we have
previously demonstrated). Assuming otherwise will add an additional
term ($t_\text{repro}$) common to both signals, that will cancel in
$\tau$.

\begin{figure}[h]
  \centering
  \includegraphics[scale=0.47]{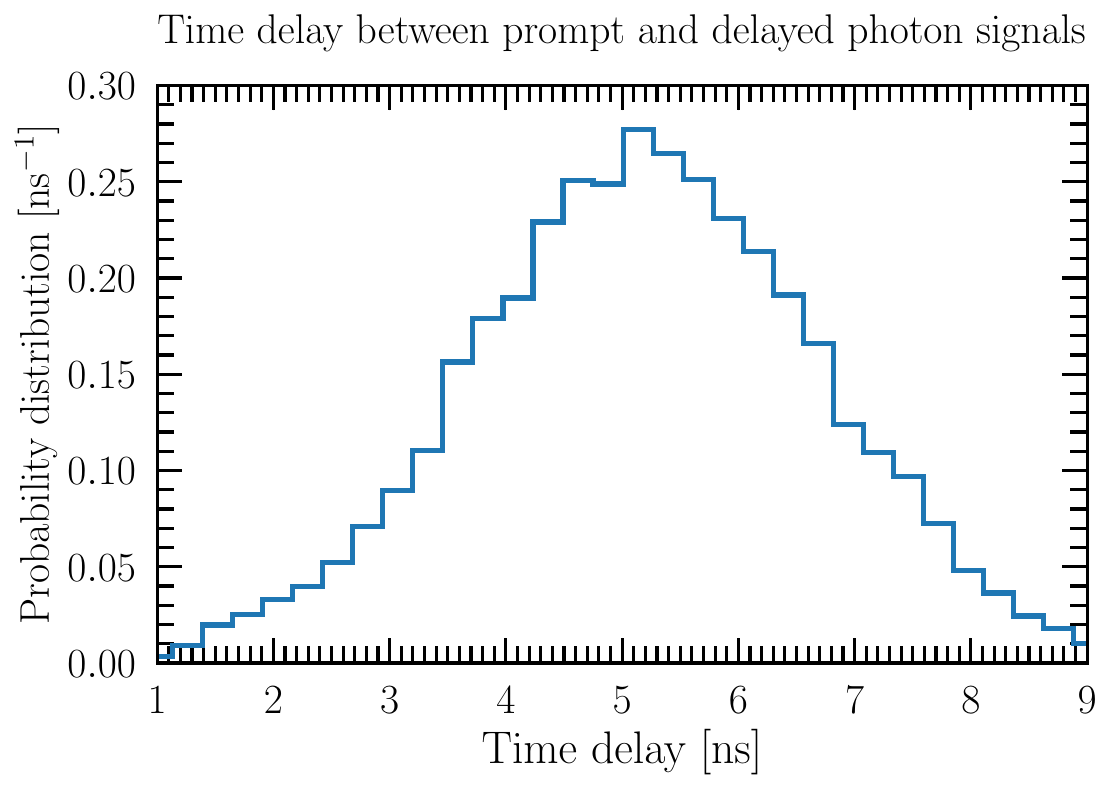}
  \caption{Time delay for final-state photon signals and scintillation
    photons produced in nuclear recoil processes. Our main assumption
    here is that final-state photons reach the SiPM array after the
    interaction, with no reprocessing by the LAB cocktail (see text
    for details).}
  \label{fig:time_delay_P}
\end{figure}

For the calculation of $t_\text{delayed}$ we have assumed
$\tau_\text{recoil}\ll \tau_\text{decay}$, with
$\tau_\text{recoil}\subset [0,1]\,$ns. For $\tau_\text{decay}$ we
assume a Gaussian distribution with mean located at
$\overline{\tau}_\text{decay}=4.69\,$ns and standard deviation
$\sigma=1.5\,$ns \cite{Lombardi:2013nla}. Results are shown in
Fig. \ref{fig:time_delay_P}, where it can be seen that the average
time window amounts to about 5 ns.

\begin{figure*}[t]
  \centering
  \includegraphics[scale=0.47]{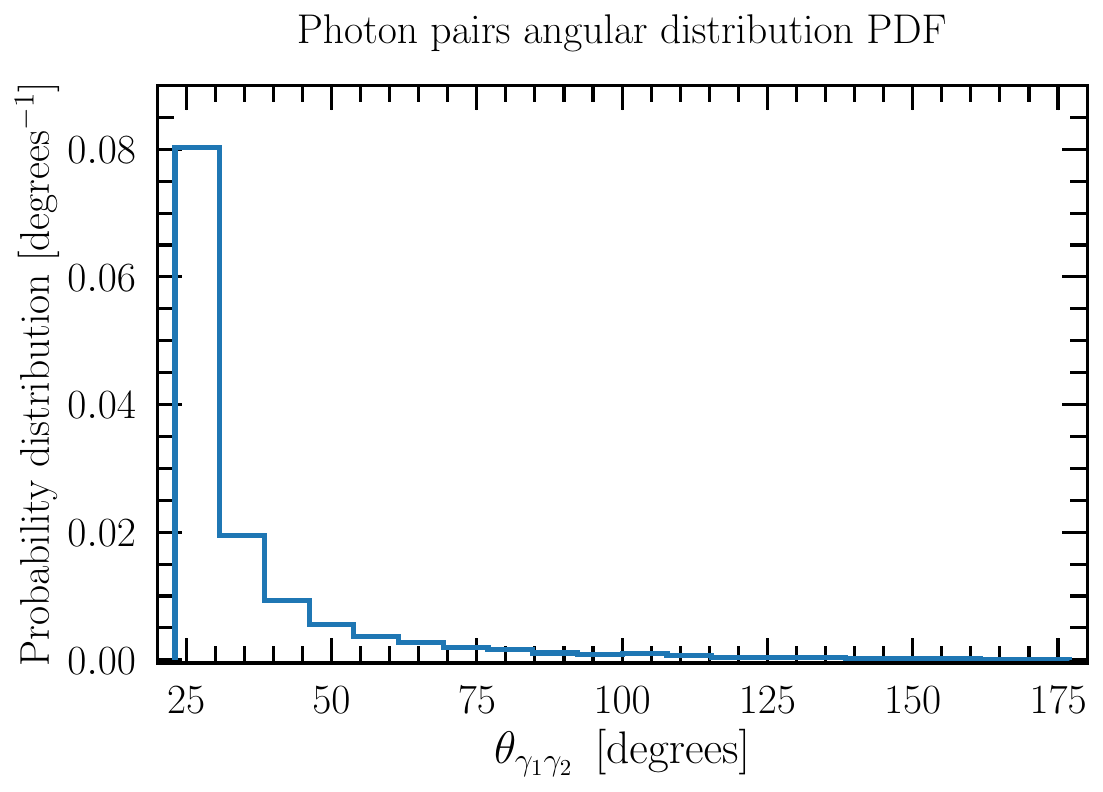}
  \includegraphics[scale=0.47]{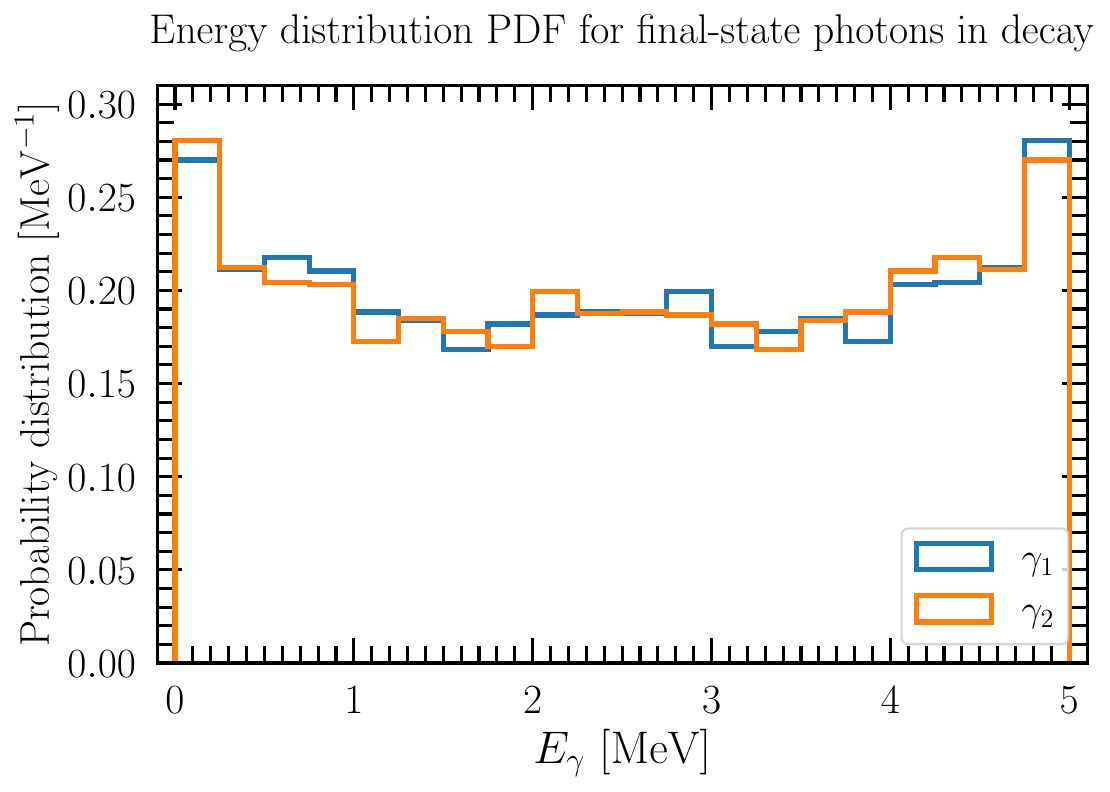}
  \caption{\textbf{Left graph}: PDF for the relative direction between
    final-state photons produced in ALP decays and measured in the
    laboratory frame. \textbf{Right graph}: Final-state photons energy
    distribution PDF. For both calculations the kinematic point
    $E_a=5\,$MeV and $m_a=1\,$MeV has been chosen.}
  \label{fig:decays_gagg_thetagg_and_Edistribution_PDF}
\end{figure*}
All in all, it seems to us that ALP signals through Primakoff
processes might be identifiable by using the delay between the two
types of light signals they involve. For this to be the case, the SiPM
array should be able to separate signals within a $\sim 5\,$ns window
\footnote{As far as we understand, this time window will be unlikely
  to be reached with the current JUNO-TAO configuration. In contrast, with
  the more sophisticated CLOUD detector, such a discrimination may be
  obtained. Still, we believe interesting to develop here a first
  hypothetical computation for our generic detector and perform a
  specific study of this topic in future work.}. Furthermore, the
light yield in nuclear recoils should be intense enough. If these two
requirements are met, identification appears to be feasible (at least
in certain regions of parameter space).

\textit{ALP decay through $g_{a\gamma\gamma}$} produces no nuclear
recoil. In the ALP decay-rest-frame (RF), two back-to-back photons
with $E_{\gamma_1}^*=E_{\gamma_2}^*=E_\gamma^*=m_a/2$ are produced
($\gamma_1$ and $\gamma_2$ refer to the final-state photons and
$E_\gamma^*$ to energy measured in the RF). The signal, however, is
observed in the laboratory frame (LF). Thus, in order to determine
their spectrum we assume---in full generality---that in the RF
\begin{equation}
  \label{eq:momentum_p1_RF}
  \vec{p}_1=m_a(\sin\theta^*\sin\varphi^*,\sin\theta^*\cos\varphi^*,\cos\theta^*)/2\ ,
\end{equation}
with $\vec{p_1}=-\vec{p_2}$ (as dictated by momentum
conservation). For an ALP propagating along a unitary vector
$\widehat{n}=(\sin\theta\sin\varphi,\sin\theta\cos\varphi,\cos\theta)$
in the laboratory frame (LF), a general Lorentz transformation enables
writing
\begin{align}
  \label{eq:boosted_energy}
  E_{\gamma_1}&=\frac{m_a}{2}\gamma\left[1 + \vec{\beta}\cdot\widehat{n}\right]\ ,
                \nonumber\\
  E_{\gamma_2}&=\frac{m_a}{2}\gamma\left[1 - \vec{\beta}\cdot\widehat{n}\right]\ ,
\end{align}
with the boost and velocity factors given by $\gamma=E_a/m_a$ and
$|\vec{\beta}|=\sqrt{\gamma^2-1}/\gamma$. Note that
Eq. (\ref{eq:boosted_energy}) is inline with expectations:
$E_{\gamma_1}+E_{\gamma_2}=m_a\gamma/2$. And that
$E_{\gamma_i}\subset
\gamma\;m_a\left[1-|\vec{\beta}|,1+|\vec{\beta}|\right]/2$, subject to
energy conservation.

\begin{figure*}[t]
  \centering%
  \includegraphics[scale=0.47]{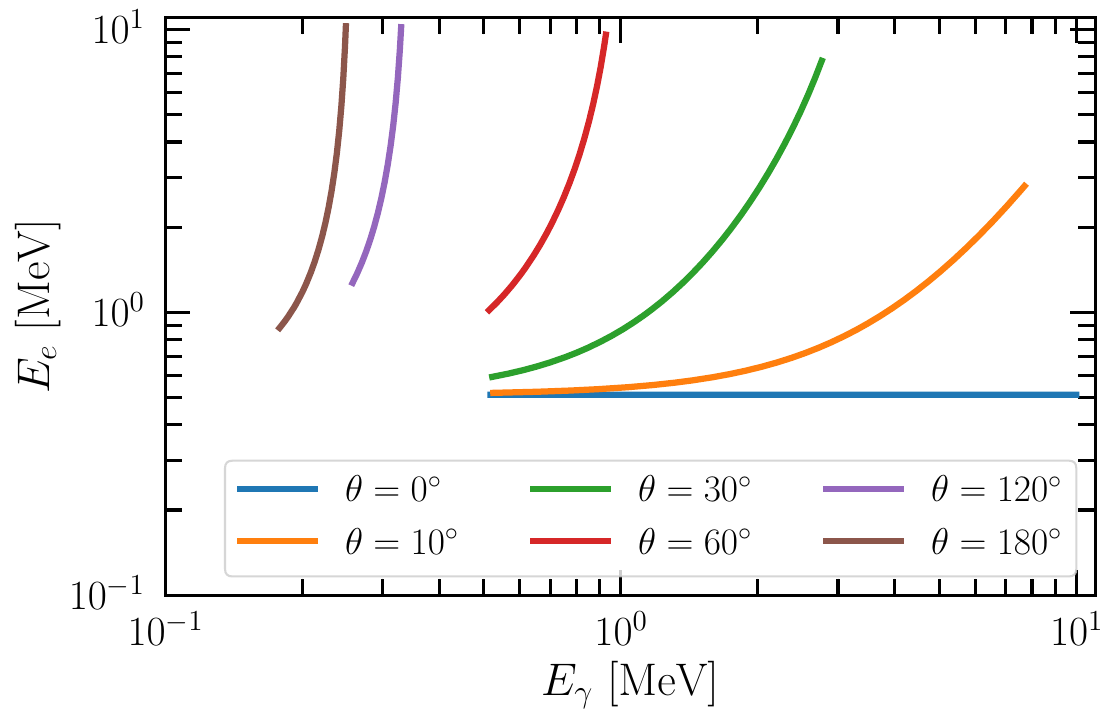}
  \includegraphics[scale=0.47]{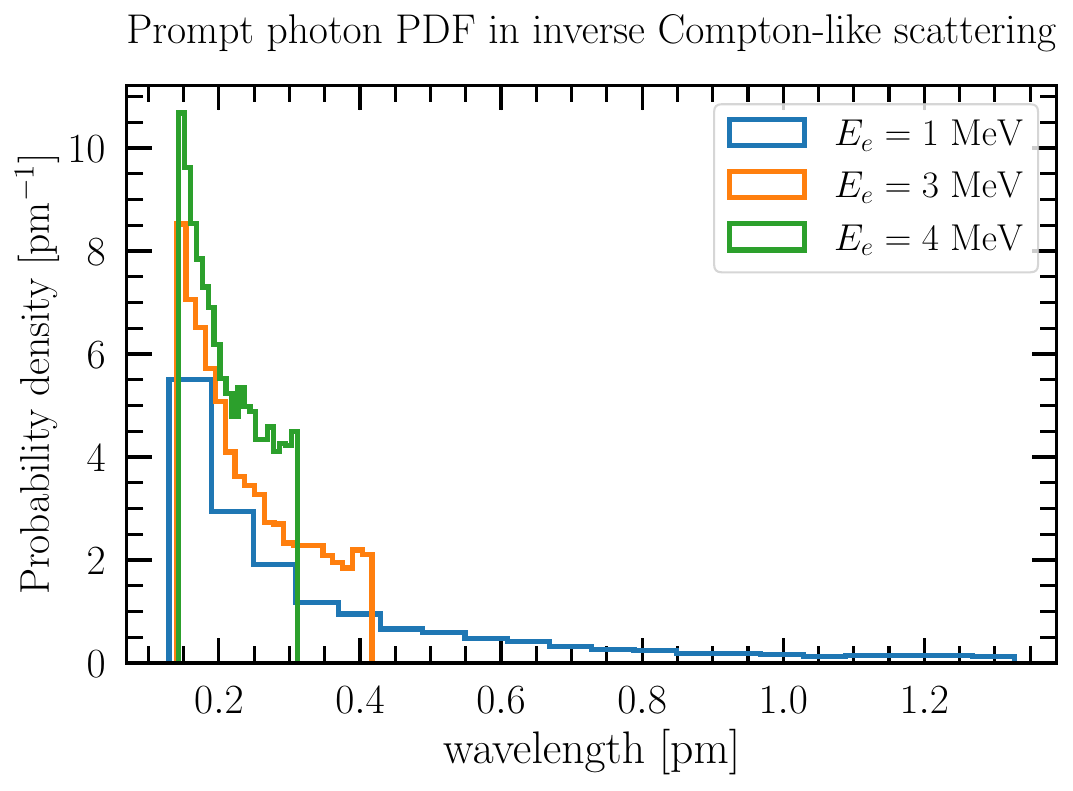}
  \caption{\textbf{Left graph}: Final-state photon and electron
    energies in inverse Compton-like scattering calculated for
    $m_a=100\,$keV and different ALP and photon directions
    ($\theta$). \textbf{Right graph}: Prompt photon kinematic spectrum
    (final-state photon) in inverse Compton-like scattering, assuming
    forward final-state electron (backward photon). The same ALP mass
    as in the left graph has been assumed.}
  \label{fig:Eg_Ee_Compton_like}
\end{figure*}
To determine the degree at which the photons are still back-to-back in
the LF, one could rely on the photons 3-momentum components written in
the LF. These expressions, however, are a complicated function of the
angular variables. So we instead restore to 4-momentum conservation in
the LF, from which it follows that
\begin{equation}
  \label{eq:relative_direction_g1_g2}
  \cos\theta_{{\gamma_1}{\gamma_2}}= 1 - \frac{m_a^2}{2E_{\gamma_1}E_{\gamma_2}}\ .
\end{equation}
Eq. (\ref{eq:relative_direction_g1_g2}), along with those in
Eq. (\ref{eq:boosted_energy}), enable the determination of the photons
relative direction PDF as well as the photons kinematic spectrum
PDFs. To illustrate their behavior, we have sampled over 10000 events,
assuming $E_a=5\,$MeV and $m_a=1\,$MeV. Results are shown in
Fig.~\ref{fig:decays_gagg_thetagg_and_Edistribution_PDF}.

In contrast to what happens in the RF, almost all events final-state
photons are rather collinear\footnote{This is just a manifestation of
  the relativistic beaming effect, whereof radiation emitted by a
  relativistic source is enclosed within a cone of half-angle
  $1/\gamma$. The more boosted the particle is, the more collinear the
  emitted photons will be.}. Since in the laboratory frame
$\vec{p}_a=\vec{p}_{\gamma_1}+\vec{p}_{\gamma_2}$, with
$\vec{p}_a\neq \vec{0}$, this is somewhat expected. The kinematic
spectrum distributions reveal another interesting feature. Although
not very pronounced, events tend to cluster at the extremes of the
allowed energy window. Thus, for those $\gamma_1$ clustering in the
upper part of the interval, the same amount of $\gamma_2$ will cluster
in the lower part of that range (and vice versa). So, ALP decays will
produce two almost collinear photons, one carrying most of the
available energy ($m_a\times\gamma$) and the other one only a small
fraction. There are of course other configurations in which the
available energy is more uniformly distributed, but according to the
kinematic spectrum PDFs those events are less likely (although not
drastically).
\subsubsection{Inverse Compton-like scattering, axio-electric
  absorption and ALP decay to electron pairs}
\label{sec:inverse_compton_like}
We now turn to the case of ALP-electron (\textit{inverse
  Compton-like}) scattering. In many aspects, signals resemble those
found in the inverse Primakoff case but with an enhancement in recoil
energies, which favors signatures identification. The final-state
photon energy follows from Eq.\,(\ref{eq:Egamma_inverse_Primakoff}) by
trading $m_N\to m_e$. Because of $E_a> m_e$, the process can produce
either forward ($E_\gamma^\text{max}$) or backward photons
($E_\gamma^\text{min}$), in stark contrast to Primakoff
scattering.
Therefore, for forward (backward) photons, the final-state electron energy will
reach a minimum (maximum), $E_{e_{\rm{min,max}}} = E_a - E_\gamma^{\text{max,min}}+m_e$.
  Fig. \ref{fig:Eg_Ee_Compton_like} (left graph) shows the
behavior in terms of the direction of the incoming ALP and outgoing
photon. Typical energies are of the order of MeV, but large photon
energy values imply lower electron energies and vice versa.

Scintillation light from electron recoils produces then much more PEs
than those produced in nuclear recoils. It can readily produce 4500
PEs, or even more depending on the relative direction between the ALP
and the photon. The shape of the scintillation light spectrum (delayed
signal) follows the Gaussian distribution in
Fig. \ref{fig:emission_spectrum}, and so the PDF for electron recoil
scintillation light emission is expected to follow that in
Fig. \ref{fig:PDF_delayed_prompt_Primakoff} (left graph).

For the prompt photon kinematic spectrum PDF, we have relied on
momentum conservation which relates the initial-state ALP energy with
that of the final-state electron and photon in terms of the ALP and
electron direction ($\alpha$).  An example of the PDF is shown in
Fig. \ref{fig:Eg_Ee_Compton_like} (right graph), assuming forward
electron events (backward photons). This result is, of course,
representative of only that kinematic configuration and different
values for $\alpha$ will produce different spectra. However, it shows
that a full characterization of the prompt signal might be possible.

The time delay between the scintillation light produced in electron
recoils and the final-state photons is not expected to change compared
with that found in the Primakoff case. The conditions under which it
was derived do not seem to depend on the recoiling particle. Thus, the
combination of prompt and delayed signals---along with a larger number
of PEs---indicate that detection of ALPs coupled to electrons is much
more favorable than detection of ALPs coupled to light.

\textit{Axio-electric absorption} will happen mainly on carbon atoms,
as it is the case for inverse Primakoff and Compton-like
scattering. Carbon atomic electron configuration is
$1\text{s}^22\text{s}^22\text{p}^2$: K-shell, (full filled 1s
subshell) and L-shell (full filled 2s subshell and partially filled 2p
subshell). In the process, an ALP gets absorbed, and an electron is
ejected. Its kinetic energy is given by
\begin{equation}
  \label{eq:axio_electric_absoption_energy}
  T_e = E_a - B\ ,
\end{equation}
where $B$ is the binding energy of the ejected electron.  Given the
average energy of the initial-state ALPs, ejected electrons will come
from the K-shell. For a free carbon atom, the K-shell binding energy
is $\sim 284.3\,$eV \cite{NIST_spectroscopy}, which means that
$T_e\simeq E_a$. After ejection, an L-shell electron will fill the
K-shell vacancy. In carbon---and in general, in low $Z$ atoms---the
likelihood for photon emission is low, so atomic reconfiguration
produces an Auger electron instead~\cite{Berkowitz:1979}: The L-shell
to K-shell transition will transfer energy to another L-shell electron
that will be ejected. The energy of the Auger electron can then be
written as
\begin{equation}
  \label{eq:Auger_electron}
  E_\text{Auger} = E_\text{K} - E_{\text{L}_1} - E_{\text{L}_2}\ ,
\end{equation}
where $E_\text{K}$ is the K-shell binding energy and $E_{L_1}$ and
$E_{L_2}$ are the binding energies for 2s and 2p electrons,
respectively. Their values---taken from
Ref. \cite{NIST_spectroscopy}---are $E_{L_1}\simeq 18.10\,$eV and
$E_{L_2}\simeq 11.30\,$eV, so $E_\text{Auger}=254.9\,$eV. All in all,
axio-electric absorption produces first an electron with energy
 $E_e=m_e+T_e \approx m_e+E_a$ , followed by an Auger electron with a well
determined energy.

Thus, in this case as well, one can properly define a prompt and a
delayed signal with a time window determined by the time elapsed
between axio-electron emission and Auger emission. Experimental
studies of those time windows for neon, using intense X-ray
free-electron lasers, have shown $\tau_\text{Auger}=2.2\,$fs
\cite{Haynes_2021}. The order of magnitude for the time window in this
case is, therefore, $1\,$fs. There are, of course, some caveats to this
discussion, of which we would like to mention at least two. First of
all, Auger electrons are very soft electrons and so once emitted they
might be absorbed by the medium. If this is the case, they will not
leave a measurable signature. Secondly, for the prompt and delayed
signals to be of any use, the SiPMs should have time resolutions in the
fs range. For that reason, we only want to stress that if the
conditions we have outlined could somewhat be met, the possible
background that axio-electric absorption might face could perhaps be
largely mitigated.

We now turn to the final process that ALP-electron couplings enable,
namely, \textit{ALP decay to $e^+e^-$ pairs}. Differences with decay
to final-state photons arise only because of the massive final
states. The electron and positron energies are dictated by expressions
like those in Eq.~(\ref{eq:boosted_energy}), with the last term traded
according to
$\vec{\beta}\cdot\widehat{n}\to \vec{\beta}\cdot \vec{v}_e$
($|\vec{v}_e|=\sqrt{1-4m_e^2/m_a^2}$ the electron/positron
velocity). The electron and positron energies are then in the MeV
range. Accordingly, their propagation in the detector medium will
produce an abundant amount of scintillation light. In the LAB frame,
their relative direction $\theta_{e^-e^+}$ is given by
\begin{equation}
  \label{eq:costhetae+e-}
  \cos\theta_{e^-e^+}\simeq 1 - \frac{m_a^2-2m_e^2}{2E_{e^+}E_{e^-}}\ ,
\end{equation}
where we have used the fact that---in general---the electron-positron
pair is very boosted and so $|p_{e^\pm}|\gg m_e$. Compared with
Eq.~(\ref{eq:relative_direction_g1_g2}), one would expect the opening
angle to be smaller than in the final-state photon case. This,
however, is not the case because of the kinematic constraint
$m_a>2m_e$. So, regions where the electron mass could produce a
sizable suppression of the second term in Eq.~(\ref{eq:costhetae+e-})
are not kinematically accessible. Thus, pairs produced in ALP decays
are expected to be as well rather collinear. A fraction---likely
small---will reach the SiPMs, and the others will produce an abundant
amount of photons (scintillation light). Electrons will generate
scintillation light, and positrons---instead---will annihilate with
electrons, producing in doing so a broad spectrum around the 511 keV
line.
\section{Sensitivities}
\label{sec:sensitivities}
\begin{figure}
  \centering%
  \includegraphics[scale=0.47]{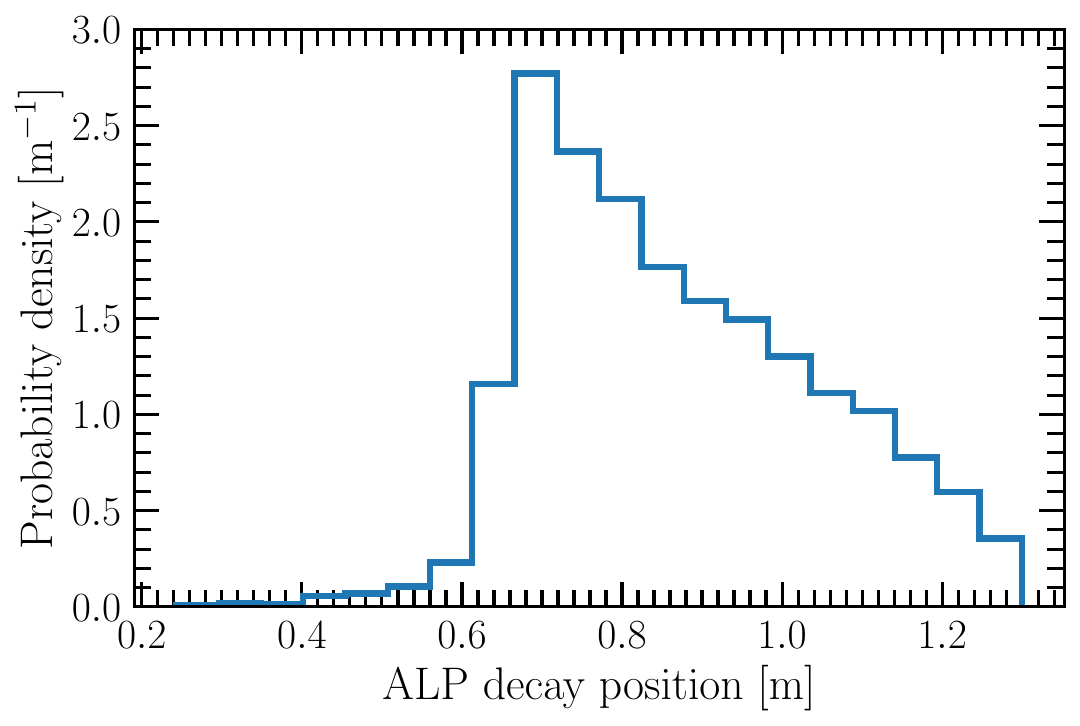}
  \caption{ALP decay position (within the fiducial volume) probability
    density function as a function of decay position. The most likely
    point is found at $r=0.7\,$m. For ALP decay event rates, we fix
    the detector depth to that value.}
  \label{fig:most_likely_decay_point}
\end{figure}

\begin{figure*}[t]
  \centering
  \includegraphics[scale=0.48]{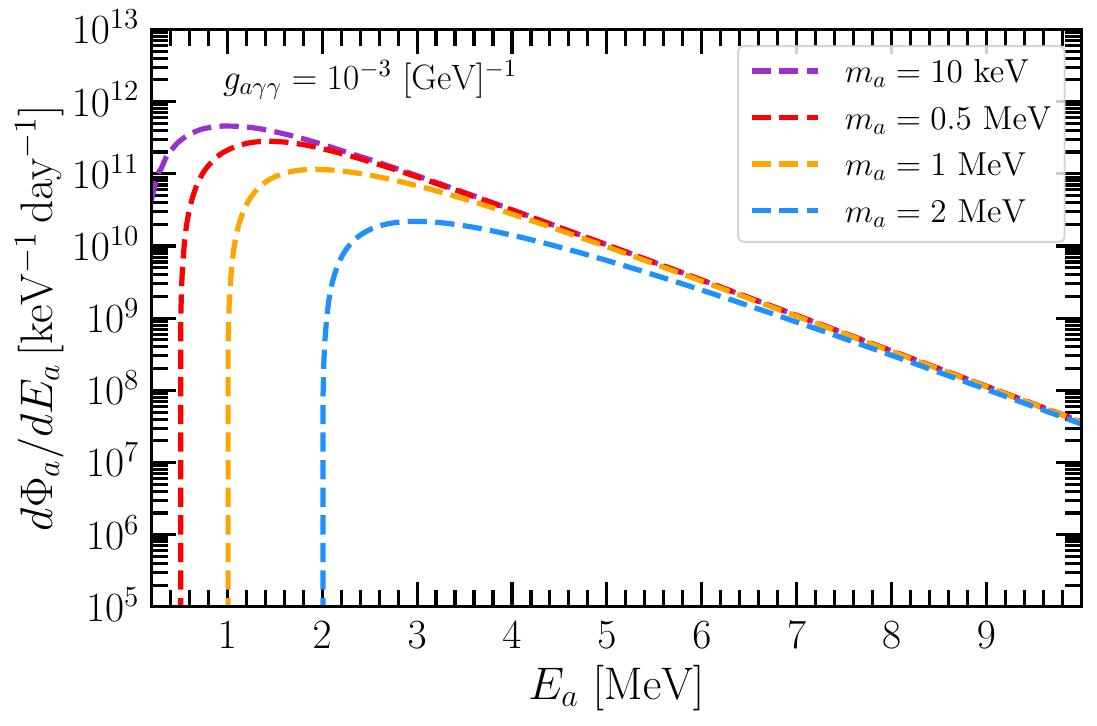}
  \includegraphics[scale=0.48]{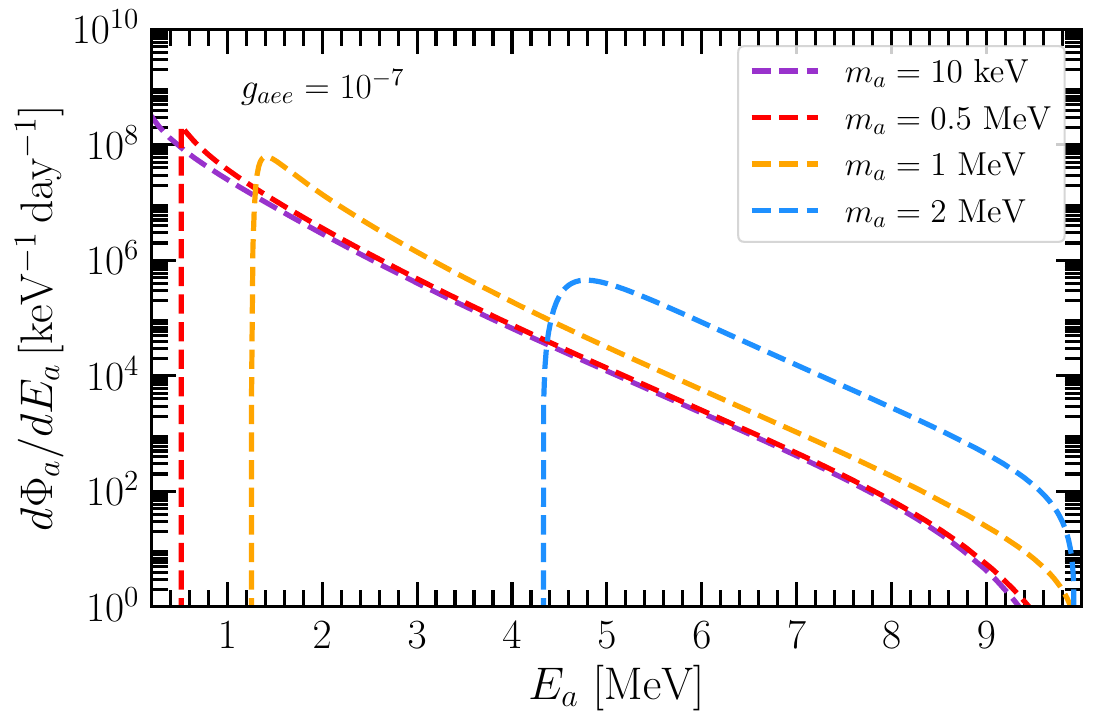}
  \includegraphics[scale=0.46]{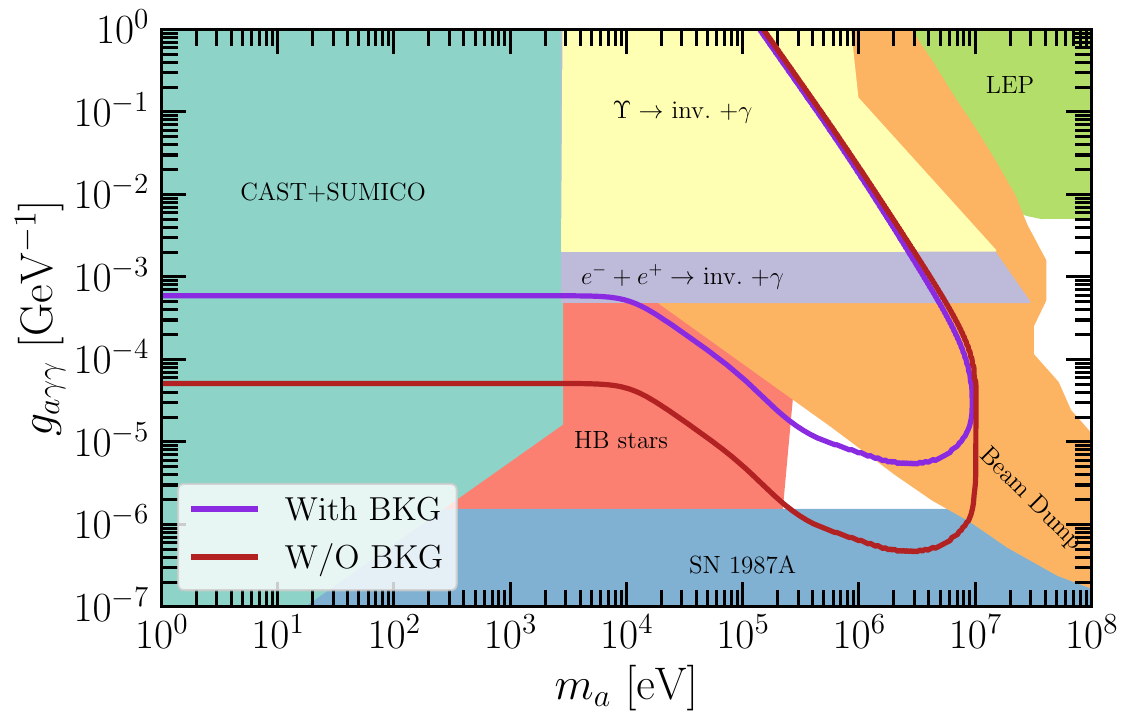}
  \includegraphics[scale=0.46]{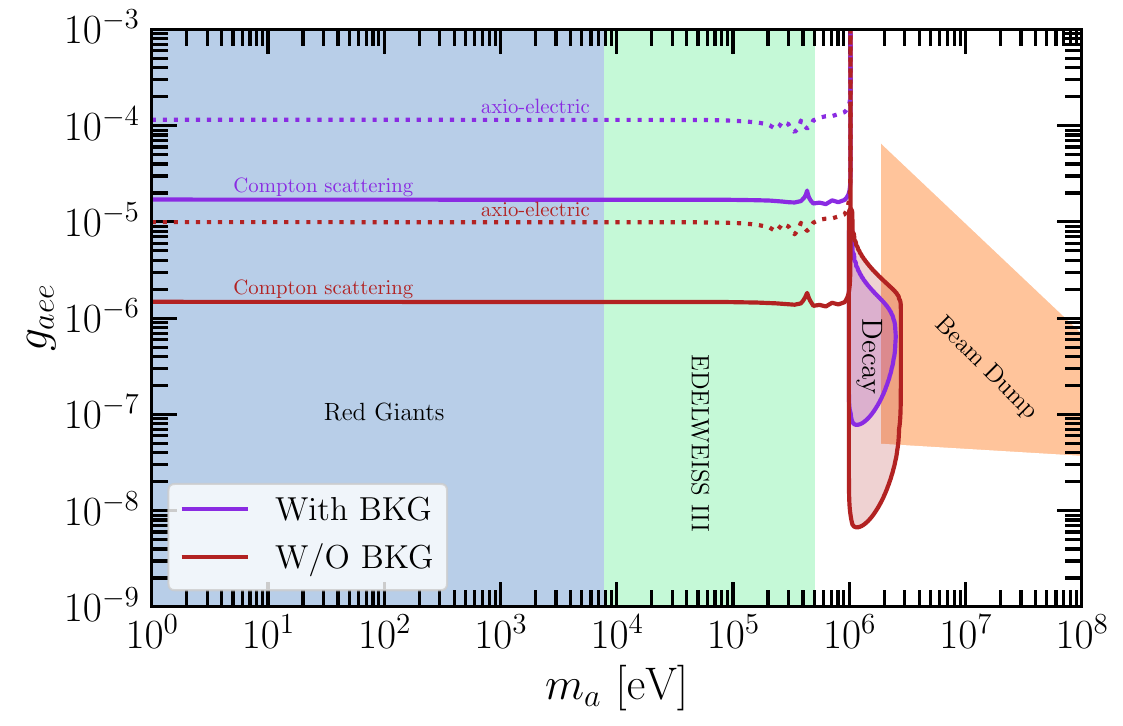}
  \caption{\textbf{Upper left graph}: Expected ALP fluxes from
    Primakoff scattering processes as a function of ALP energies for
    different ALP masses and
    $g_{a\gamma\gamma}=10^{-3}\,\text{GeV}^{-1}$. \textbf{Upper right
      graph}: Same as upper left graph but for Compton-like scattering
    processes. \textbf{Lower left graph}: 90\% CL sensitivities,
    obtained by assuming an exposure of 180 days, shown in the
    $g_{a\gamma\gamma}-m_a$ plane for two extreme scenarios: (i) Full
    background (from singles from radioactivity \cite{JUNO:2020ijm})
    plus signal, (ii) background-free searches.  \textbf{Lower right
      graph}: Same as lower left graph, but in the $g_{aee}-m_a$
    plane.}
  \label{fig:ALP_fluxes}
\end{figure*}
In this section, we discuss ALP search sensitivities at the JUNO-TAO
detector. For that aim, we rely on the results presented in
Secs. \ref{sec:ALP_prod} and \ref{sec:ALP_signals}. In particular, we
entertain the possibility that because of the prompt and delayed
photon signals, as well as the distinctive imprints left by the ALP
decay by-products, searches may be potentially background-free.

ALP event rates in LAB are mainly determined by carbon nuclei, as can
be seen from the hydrogen and carbon mass fractions
\begin{equation}
  \label{eq:LAB_mass_fractions}
  f_\text{C}^n=\frac{m_\text{C}(n+6)}{m_\text{LAB}}\ ,
  \qquad
  f_\text{H}^n=\frac{m_\text{H}(2n+6)}{m_\text{LAB}}\ .
\end{equation}
Here, we assume $\text{C}_6\text{H}_5\text{C}_{13}\text{H}_{27}$ as a
material target while neglecting all the other components of the
JUNO-TAO cocktail. An assumption well justified given their relative
small amounts. For this compound, we find that carbon contributes to
$88\%$ of the total mass. Relevant for the event rate calculation as
well are the nuclear reactor power, the baseline, and detector radius,
which we fix to $P=4.6\;$GW, $L=30\;$m, and
$R_\text{JUNO-TAO}=0.65\;$m (active volume radius).

For event rates from ALP decays, the detector depth, $r$, is also
required [see Eq. (\ref{eq:Prob_decay})]. Given the geometry, we
proceed by noting that the detector will be deployed at the Taishan
Neutrino Laboratory located in a basement at 9.6 m underground
\cite{JUNO:2020ijm}. With that information, along with the detector
fiducial volume radius and baseline, a simple two-dimensional model of
ALP decay within the fiducial volume can be constructed (see
App. \ref{sec:model_ALP_int_FV}). That model enables the determination
of the most likely point at which decay will
happen. Fig. \ref{fig:most_likely_decay_point} shows the probability
distribution function in terms of the decay point length. This result
shows that decays happen more often at $r\subset [0.66,0.72]\,$m, and
so we fix $r=0.7\,$m in our calculation.

The expected qualitative behavior of sensitivities can be understood
already from the results presented in
Sec. \ref{sec:ALP_prod}. Fig.~\ref{fig:ALP_fluxes} (upper panels)
shows results for the predicted ALP fluxes in terms of ALP
energies. First of all, since at low ALP masses the differential cross
sections have little dependence upon $m_a$, ALP fluxes are degenerate
regardless of the production mechanism. At high ALP energies, close to
threshold, fluxes tend to degenerate as well. An effect attributed to
the fact that as energy increases terms that involve ALP masses become
subdominant.

For the Primakoff and inverse Primakoff case, this implies that for
$m_a\lesssim 10\,$keV sensitivities depend only on
$g_{a\gamma\gamma}$. For masses up to $100\,$keV, terms involving ALP
masses in the scattering cross-section are still of little
relevance. However, at those masses, the decay width reaches values
such that the decay probability becomes sizable enough [the decay
width scales as $m_a^3$, see Eq. (\ref{eq:total_width_agg})]. Thus,
event rates will become sensitive to ALP masses with increasing ALP
mass.  Scattering processes will also introduce a mass dependence,
which will be mild due to the decreasing fluxes.  The strongest
dependence on ALP mass is thus expected from ALP decays.

For processes controlled by the $g_{aee}$ coupling, sensitivities are
expected to be ALP mass independent up to larger ALP masses. In the
low-mass region fluxes are rather degenerate. With increasing ALP
mass, that degeneracy is to a certain extent lifted, but with
decreasing fluxes (see Fig. \ref{fig:ALP_fluxes}, right upper
graph). Thus, as in the case of the $g_{a\gamma\gamma}$ coupling, the
most pronounced ALP dependence is introduced by decays. In this
case---however---the decay becomes effective only when
$m_a>1\;\text{MeV}$ value at which there is no flux for a large enough
coupling. For an ALP to reach the detector, its decay length should
exceed the reactor-detector baseline, implying the constraint
\begin{equation}
  \label{eq:decay_length_gaee}
  g_{aee} \lesssim \frac{8\pi\,\hbar\,c}
  {L\,m_a\,\sqrt{1-4m_e^2/m_a^2}}\ .
\end{equation}
This limit is satisfied for any ALP mass in the kinematically allowed
range provided $g_{aee}\lesssim 10^{-7}$.

With the qualitative behavior discussed, we then determine
sensitivities with the aid of the following $\chi^2$ function
\begin{equation}
  \label{eq:chi_square}
  \chi^2(g_X,m_a)=\frac{N^2_\text{signal}(g_X,m_a)}
  {N_\text{signal}(g_X,m_a)
    + N_\text{bckg}
    + N^2_\text{signal}\sigma_\text{sig}^2}\ .
\end{equation}
Here $N_\text{signal}$ is the number of ALP events for a point in
parameter space $\{g_X,m_a\}$ ($g_X=g_{a\gamma\gamma}$ or
$g_X=g_{aee}$), $N_\text{bckg}$ is the number of background events,
and $\sigma_\text{sig}^2$ the signal systematic uncertainty. Note that
when writing the $\chi^2$ function as in Eq. (\ref{eq:chi_square}) we
are assuming that the background systematic uncertainty,
$\sigma_\text{bckg}$, amounts to zero. This is certainly not the case,
but we have checked that the signal systematic uncertainty has little
impact and so found no reason for the background systematic
uncertainty to do so. We have fixed $\sigma_\text{sig}=0.1$.

Results for 90\% CL sensitivities (obtained by the condition
$\chi^2(g_X,m_a) = 4.61$ for 2 degrees of freedom) are shown in the
lower graphs of Fig. \ref{fig:ALP_fluxes}.  Lower left panel in the
$g_{a\gamma\gamma}-m_a$ plane, while the lower right panel in the
$g_{aee}-m_a$ plane.  Results are presented for two extreme search
scenarios: (i) Signal+background and (ii) signal alone
(background-free). At JUNO-TAO, ALP searches are subject to an
abundant background, which includes: IBD, fast neutrons after veto,
and singles from radioactivity (see
e.g. Refs. \cite{JUNO:2020ijm,Xu:2022mdi} for details). Particularly
threatening are the latter, which are overwhelmingly larger than those
from IBD or fast neutrons, amounting to about $10^9$ events/year.

Sensitivities in the signal+background scenario---calculated for a
six-month data-taken period---thus include only the singles from
radioactivity background, in line with the analysis presented in
Ref. \cite{Smirnov:2021wgi}. The background-free scenario is motivated
by results from Sec. \ref{sec:ALP_signals} which show: Presence of
prompt and delayed photon signals in scattering processes, coincident
events in the case of ALP decays, and prompt (scintillation light) and
delayed signals (Auger electrons) from axio-electric absorption. When
presenting results for this scenario, we do not aim at suggesting that
such an ideal case can be experimentally achieved. Instead our idea is
that of showing that actual results might---in principle---interpolate
between the worst-case (signal+background) and best-case (background-free) scenarios.

In the case of the nuclear channel, sensitivities in scenario (i)
indicate that a certain portion of the cosmological triangle can be
explored. This region is subject to constraints from Big Bang
nucleosynthesis and from the effective number of relativistic species
\cite{Cadamuro:2011fd,Millea:2015qra,Depta:2020wmr}. In terms of the
$\Lambda\text{CDM}$ model, where inflation is assumed to have happened
at very early times, it is therefore ruled out. It opens up, however,
in e.g. low-reheating models \cite{Depta:2020wmr}. Testing this region
in a controlled experimental environment, therefore, provides an
indirect test for nonstandard cosmological scenarios. Reducing the
background will improve this result up to the point where the whole
cosmological triangle can be thoroughly covered, provided
background-free searches can be carried out with the aid of the prompt
and delayed photon signals we have identified. It will enable as well
to enter the region where horizontal branch (HB) stars and supernova
(SN) constraints apply. Thus, it will test ALP models where medium
effects forbid their production in astrophysical environments (see
e.g. Ref.~\cite{Jaeckel:2006xm}).

Results in Fig. \ref{fig:ALP_fluxes} (lower left graph) include as
well 90\% CL constraints from other astrophysical and laboratory
experiments. CAST and SUMICO solar axion searches
\cite{CAST:2008ixs,Minowa:1998sj}, HB stars, and SN1987A
\cite{Ayala:2014pea,Carenza:2020zil,Lucente:2020whw}. The latter
derived from stellar cooling and SN energy loss arguments. Laboratory
searches include rare $\Upsilon$ invisible decays, exotic final states
in $e^+e^-$ collisions (including LEP) and beam dump searches
\cite{CrystalBall:1990xec,BaBar:2008aby,Hearty:1989pq,
  Riordan:1987aw,Bjorken:1988as,Dobrich:2017gcm}

Turning to the case of the electron channel, sensitivities can highly
improve in the background-free case as well. Results in the
signal+background scenario show that in regions of MeV ALP masses,
sensitivities go down to couplings of the order of
$10^{-7}$. Background-free searches will be able to scan couplings
more than one order of magnitude smaller. Interestingly, as far as we
know, this region is unexplored. So, if the background can be
efficiently mitigated this region might be thoroughly explored. Taking
advantage of coincident events from ALP decays, discussed in
Sec. \ref{sec:ALP_signals}, might help in doing so. Furthermore,
models where environmental effects are at work can be tested in those
regions where Red Giants constraints apply \cite{Raffelt:2006cw}. Note
that further limits, in this case, include searches in the
EDELWEISS-III array of germanium bolometers \cite{EDELWEISS:2018tde}
and beam dump laboratory searches \cite{Essig:2010gu}.
\section{Conclusions}
\label{sec:conclusions}
In this paper, we have studied the capability of a JUNO-TAO-like
detector concept as a case study of ALP searches in short-baseline
neutrino experiments using large organic liquid scintillator
detectors. To do so, we have considered ALP couplings to photons and
electrons. In the former case, our analysis includes production
through Primakoff-like scattering processes and detection through its
inverse equivalent and decay to photon pairs. In the latter, instead,
production induced by Compton-like scattering processes and detection
through all the possible channels the coupling allows for (apart from
$e^+e^-$ pair production in the electron field): Inverse equivalent
process, axio-electric absorption, and decay to $e^+e^-$ pairs.

We have shown that signals in both nuclear and electron recoil
processes produce prompt and delayed photon signals. The prompt
component arises from the final-state photon produced in the
scattering process itself. The delayed signal, instead, from
scintillation light produced by the recoil. The time window, estimated
to be of the order of 5 ns (for the experimental parameter
configuration considered), has its origin in the time it takes the
scintillation light to be built. We have pointed out that if the SiPMs
can resolve this time window, an efficient radioactive background
rejection might be possible. Although we have considered a generic
liquid scintillator detector, when comparing with detectors currently
under construction, we find that the CLOUD experiment may be more
adequate for exploring this time window and deserves a more dedicated
study. In such scenario, we have demonstrated that sensitivities can
improve largely. If this turns out to be the case, a full exploration
of the so-called cosmological triangle seems feasible. Furthermore, in
the case of ALPs coupled to electrons it will improve sensitivities by
one order of magnitude allowing for searches in regions not yet
explored.

\begin{figure*}[t]
  \centering
  \includegraphics[scale=0.9]{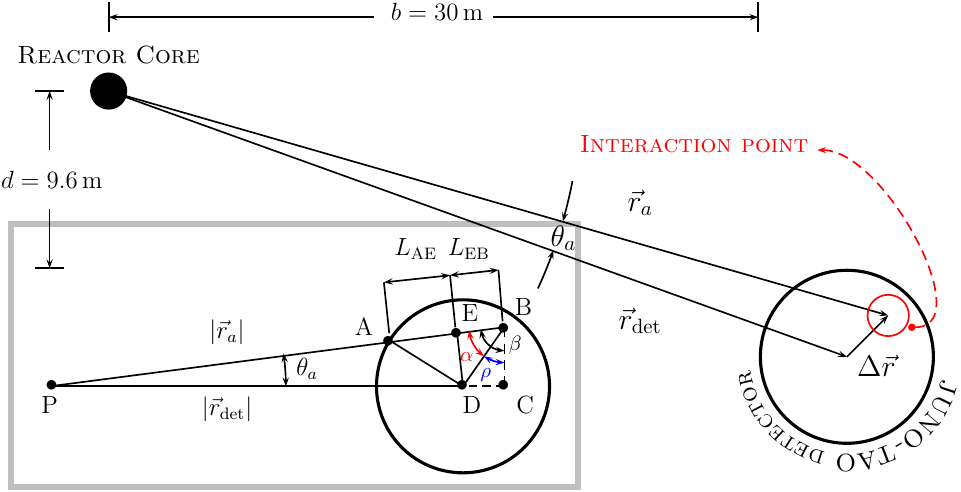}
  \caption{Not-to-scale sketch of the two reactor cores of the Taishan
    Nuclear Power Plant and the JUNO-TAO detector. ALP trajectory is
    determined by $\vec{r}_a$, while the interaction point within the
    detector fiducial volume by
    $\Delta \vec{r}=\vec{r}_a-\vec{r}_\text{det}$. Baseline,
    $b=30\;$m, as well as detector depth, $d=9.6\,$m, are taken from
    Ref. \cite{JUNO:2020ijm}. This sketch is employed in the
    determination of the most likely ALP decay location. The inset
    sketch shows the geometry employed for the determination of the
    ALP most likely decay point within the fiducial volume.}
  \label{fig:sketch_detection}
\end{figure*}
In the case of axio-electric absorption, we have found as well a
prompt and a delayed component. Ejected electrons from carbon atoms
K-shells will be followed by Auger electrons in a very narrow time
window of the order of $1\;$fs. We have argued that time resolution,
along with the small amount of scintillation light produced by the
Auger electron, seems to be an experimental challenge. If these two
issues can be properly addressed, background effects may be
efficiently mitigated.

Decays will directly produce photon pairs in the case of ALPs coupled
to photons. They will produce scintillation light and a broaden
spectrum around 511 keV from $e^+e^-$ annihilation, in the case of
ALPs coupled to electrons. In both cases, the signals will be subject
to relativistic beaming effects and so will lead to coincident photon
events with small opening angle. Assuming small diffusion effects,
this will translate into activation of nearby SiPMs.

All in all, by using a JUNO-TAO-like detector concept we have argued
that short-baseline large liquid scintillator detectors are very well
suited for ALP searches in the MeV mass window. Both electron and
nuclear scattering seem to be promising channels, in particular, if
the experimental challenges we have identified can be overcome. If so,
it seems to us that this technology can provide valuable information
on the global ALP search program.
\section*{Acknowledgments}
We would like to thank Anatael Cabrera and Dimitrios Papoulias for
useful discussions and comments on this paper.  D.A.S. is supported by
ANID grant ``Fondecyt Regular'' 1221445, and H.N. by CNPq and CAPES
from Brazil.  O.G.M. would like to thank support from Sistema nacional
de investigadoras e investigadores (SNII) from Mexico. Two of us
(D.A.S. and H.N.) thank CINVESTAV for the hospitality during the
initial stage of this work.
\appendix
\section{Fiducial volume ALP decay}
\label{sec:model_ALP_int_FV}
Here we present the details of the two-dimensional model adopted for
the determination of the ALP most likely decay point. Relying on the
sketch in Fig.~\ref{fig:sketch_detection}, one can write
\begin{equation}
  \label{eq:DeltarSq}
  |\Delta \vec{r}|^2=|\vec{r}_\text{det}|^2 + |\vec{r}_a|^2 -
  2|\vec{r}_\text{det}||\vec{r}_a|\cos\theta_a\ ,
\end{equation}
where $\theta_a$ is the angle between the fixed vector
$\vec{r}_\text{det}$ and the ALP position vector measured from
the production point (assuming a point-like source) to the interaction
point. Its norm is given by $|\vec{r}_\text{det}|^2=(b+R)^2+(d+R)^2$,
where $R=0.65\;$m is the detector fiducial volume radius. For the
interaction to take place within the fiducial volume, the condition
$|\Delta \vec{r}|\subset [0,R]$ must be satisfied. With that
information at hand, Eq. (\ref{eq:DeltarSq}) can be solved for
$|\vec{r}_a|$ resulting in
\begin{equation}
  \label{eq:ra_sol}
  |\vec{r}_a|_{\pm}=|\vec{r}_\text{det}|\cos\theta_a\pm
  \sqrt{|\Delta\vec{r}|^2 - \sin^2\theta_a|\vec{r}_\text{det}|^2}\ .
\end{equation}
For this equation to be meaningful the following constraint must be
satisfied
\begin{equation}
  \label{eq:constraint}
  -\frac{|\Delta\vec{r}|}{|\vec{r}_\text{det}|}\leq \sin\theta_a\leq
  \frac{|\Delta\vec{r}|}{|\vec{r}_\text{det}|}\ .
\end{equation}
ALPs that satisfy this condition will pass by the detector. With this
condition, one can then run a simple, yet useful, Monte Carlo as
follows. Randomly sample $|\Delta \vec{r}|$ in the interval $[0,R]$
$N$ times. For each point, calculate randomly $\sin\theta_a$ in the
interval dictated by condition (\ref{eq:constraint}). The random
values $|\Delta \vec{r}|$ and $\theta_a$ obtained in this procedure
allow then to determine an interval for $|\vec{r}_a|$:
$[|\vec{r}_a|_{-},|\vec{r}_a|_{+}]$. From this interval, a random value
for $|\vec{r}_a|$ can be selected for each of the $N$ events
generated.

The information from the Monte Carlo can then be used to determine the
detector ``depth'' (point where the interaction takes place, decay in
this case), namely (see inset sketch in
Fig. \ref{fig:sketch_detection})
\begin{equation}
  \label{eq:detector_depth}
  r = L_\text{AE} + L_\text{EB}\ .
\end{equation}
First of all, from the triangle defined by vertices PBC one finds
$\beta=\pi/2-\theta_a$. From the triangle defined by vertices BCD,
instead,
\begin{equation}
  \label{eq:traingle_BCD}
  \cos\rho=\frac{L_\text{BC}}{|\Delta \vec{r}|}=
  \frac{|\vec{r}_a\sin\theta_a|}{|\Delta\vec{r}|}\ .
\end{equation}
Since $\alpha=\beta -\rho$, Eq. (\ref{eq:traingle_BCD}) along with the
value for $\beta$ thus allows writing from the triangle defined by
vertices DEB
\begin{equation}
  \label{eq:LEB}
  L_\text{EB}=|\Delta \vec{r}|\cos\alpha\ .
\end{equation}
From the same triangle, one can as well write
$L_\text{ED}=|\Delta \vec{r}|\sin\alpha$, which then, from the triangle defined by vertices DEA, one finds
\begin{equation}
  \label{eq:LAE}
  L_\text{AE}=\sqrt{R^2-L_\text{ED}^2}\ .
\end{equation}
Eqs. (\ref{eq:LEB}) and (\ref{eq:LAE}), along with
Eq. (\ref{eq:detector_depth}), define the detector depth in terms of
randomly generated quantities and so enable the construction of the
PDF shown in Fig. \ref{fig:most_likely_decay_point}.
\bibliographystyle{utphys}
\bibliography{biblio}
\end{document}